\documentclass[twocolumn]{article}
\usepackage{graphicx,color} 
\usepackage{amsmath,mathrsfs}
\usepackage{natbib}

\usepackage[english]{babel}
\usepackage[T1]{fontenc}

\newcommand{\at}{{\char '100}}
\newcommand{\am}[1]{\textcolor{blue}{#1}}

\newcommand{\DEF}{\overset{\mathrm{def}}{=}}
\newcommand{\DEFt}{\;\smash{\overset{\text{\tiny def}}{=}}}

\newcommand{\ftransfo}[1]{\mathnormal{#1}} % fontes pour les transformations  
\newcommand{\T}[1]{#1^{\mathnormal{\scriptscriptstyle T}}} 

\renewcommand{\theenumi}{(\kern -0.15ex{\roman{enumi}})}

\begin{document}
\selectlanguage{english}
\onecolumn

\title{Reflections on the four facets of symmetry: 
how physics exemplifies rational thinking}

\author{Amaury Mouchet\\ \small
Laboratoire de Math\'ematiques
  et de Physique Th\'eorique, \\ \small
Universit\'e Fran\c{c}ois Rabelais de Tours --- \textsc{\textsc{cnrs (umr 7350)}},\\ 
 F\'ed\'eration Denis Poisson,\\ \small
 Parc de Grandmont 37200  Tours,  France\\ \small
 mouchet\at phys.univ-tours.fr\\ \small
\date{\today}}

\maketitle

\begin{abstract}
In contemporary theoretical physics, the powerful notion of symmetry
stands for a web of intricate meanings among which I identify four
clusters associated with the notion of transformation, comprehension,
invariance and projection. While their interrelations are examined
closely these four facets of symmetry are scrutinised one after the
other in great detail. This decomposition allows us to
carefully examine  the multiple  different roles symmetry plays in many
places in physics. Furthermore, some connections with other disciplines
like neurobiology, epistemology, cognitive sciences and, not least,
philosophy are proposed in an attempt to show that symmetry
 can be an organising principle also in these fields.
\end{abstract}

\textsc{pacs:}  11.30.-j,      % Symmetry and conservation laws
11.30.Qc,      % Symmetry breaking
01.70.+w  % Philosophy of science


\am{A shorter version  of this manuscript \citep{Mouchet13c} was published in \textit{European Physical Journal H}. \footnote{\am{\texttt{http://dx.doi.org/10.1140/epjh/e2013-40018-4}}}
The parts in blue are additional material to this publication.}

\setcounter{tocdepth}{4}

\tableofcontents
%%%%%%%%%%%%%%%%%%%%%%%%%%%%%%%%%%%%%%%%%%%%%%%%%%%%%%%%%%%%%%%%%%%%%%%%%%%%%%%%%%%%%%%%%%%%%%
\twocolumn
\section{Introduction}\label{sec:introduction}

During the twentieth century, the development of quantum physics and
relativity came with a semantic shift of the word ``symmetry'' in
accordance with the increasing power of such a conceptual instrument.
Mathematicians and theoretical physicists have enlarged its meaning
far beyond the one it had in the previous centuries \citep{Hon/Goldstein08a}.
 Among the multiple uses of symmetry in
theoretical physics and more generally in science, I propose to
discern four clusters associated with the notion of transformation,
comprehension, invariance and projection.  After having defined these
four facets in the next section, I will discuss them individually
with more details in
sections~\ref{sec:transformations},~\ref{sec:comprehension},
\ref{sec:invariance} and~\ref{sec:projection} respectively.  These
four notions are more or less explicit and fully recognised in the
natural philosophy since its very beginning but, as far as I know,
addressed with very loose relations between them, if any.
The contemporary theoretical physics point of view on symmetry offers a
new and very precise way to tightly bind these notions together and
cross-fertilise them.  To make a long story short, originally being an
aesthetic criterion, symmetry evolves in science into an efficient method of
classification before it becomes a powerful theoretical tool to guide
and even dictate our way of constructing models and theories. More
precisely, symmetry allows
\renewcommand{\theenumi}{$\arabic{enumi}^\circ$}
\begin{enumerate}
\item\label{it:linrep} to build (linear) algebraic
  representations\footnote{\label{fn:representation}I use the epithet
    ``algebraic'' to reinforce the distinction between a) the precise
    mathematical meaning of ``representation'' in the present context
    (see also \S~\ref{subsubsec:algrep}, Fig.~\ref{fig:representation}
    below), b) the ``internal'' meaning it takes in cognitive science
    (mental representation) and c) the usual ``external'' meaning
    (model, symbol, image, spoken word, written sign, play, etc.).
    All the three significations share the common point of being
    associated to a correspondence and can refer to the target, the
    domain of a mapping, one of their elements or the mapping
    itself. Some confusion usually raises when the domain and the
    target are not clearly distinguished.};
\item\label{it:unification} to constrain yet to unify interactions in the models of theoretical physics;
\item\label{it:universal} to extract universal properties in statistical 
 physics and in non-linear dynamics; 
\item\label{it:solvdyn} to
       solve dynamical equations by reducing the numbers of degrees of
          freedom and 
          establishing the bridge between integrable and chaotic dynamics;
\item\label{it:selrule}  to predict selection rules.
\end{enumerate}
Examples of~\ref{it:unification} are the unification of inertial and
gravitational forces in the theory of general relativity and the
mixing between electromagnetic and weak interactions in the 
 Standard Model of elementary particle physics.
  Conservation laws are particularly
manifest in \ref{it:solvdyn} and \ref{it:selrule}. Above all, the
selection rules concern quantum processes but, as constraints on the
evolution of a system, they can be found in classical dynamics as
well: the conservation of the angular momentum of a particle evolving
freely within a circular billiard (with elastic reflexions on its boundaries)
 explains the impossibility of the
inversion of the sense of its rotation around the centre. But 
any small generic deformation of the billiard
boundary, even a small bump,  can revert angular momentum
(still conserving the energy) and, actually, 
 this breaking of rotational invariance gives room 
to chaotic motion.
In this
essay I will have to say more about~\ref{it:linrep} in
\S~\ref{subsubsec:algrep} and I will focus on~\ref{it:universal} in
\S~\ref{subsubsec:pruning}.

The roles \ref{it:linrep}--\ref{it:selrule} were discovered/invented
by physicists essentially during the twentieth century~\citep[for a deep insight of these
  issues see the synthesis by][in
  particular \S~1]{Brading/Castellani03a}.  These powerful functions
of symmetry, far from erasing it, strengthen what is certainly the
most important one: the fact that symmetry allows

\smallskip
\noindent$6^\circ.$ to classify.
\smallskip

 Performing classification concerns domains broader that the one of
 crystals, quantum particles, living organisms, historical periods,
 languages and philosophical camps: any law of nature can also be
 seen as a manifestation of an objective regularity and I will try to
 show how the modern interpretations of symmetry allow to give a
 precise and, above all, coherent meaning of the terms ``objective''
 and ``regularity''. It provides a rigorous ground for the
 meaning of universality. It also concerns the quantitative
 approach of inference and understanding. Some connections with the
 dual enquiry for emergence and reduction can also be drawn.  Last but
 not least, these conceptions of symmetry shed light on a (the ?)
 characteristic of intelligence, namely modeling, and on an elementary
 process necessary for rational thinking, namely abstraction.  Since,
\begin{quotation}\small{
The whole of science is nothing more than a refinement of every day
thinking. It is for this reason that the critical thinking of the
physicist cannot possibly be restricted to the examination of the
concepts of his own specific field. He cannot proceed without
considering critically a much more difficult problem, the problem of
analyzing the nature of everyday thinking} \citep[\S~1, p.~349]{Einstein36a},
\end{quotation} 
this feedback of the concept of symmetry from physics to
science and from science to rational thinking is quite natural after
all.

Twenty years ago, 
Edelman  foresaw that there may be
some possible connections between an epistemology based on norms
elaborated by natural selective processes and the notion of symmetry
as developed by physicists. In some sense, a part of the present work
provides some consistent flesh to Edelman's "vague and utopian" (his
own words) remarks: 
\begin{quotation}\small{Physics and biology will ``correspond'' with each
other in an intimate way, certainly in the next century and possibly
even sooner than that. [\dots]  Symmetry is a stunning example of how a
rationally derived mathematical argument can be applied to
descriptions of nature and lead to insights of the greatest
generality \citep[chap.20]{Edelman92a}.}
\end{quotation}

\am{In this perspective, I believe that the notion of symmetry revitalises
and clarifies the long-standing philosophical debate between realists
and their opponents.  The notions of ``reality'', ``being'',
``existence'' or even ``self'' have always been tricky because one
cannot define them without introducing some tautology.  In the endless
controversial discussions about ontology, these notions are, at best,
acknowledged as primary concepts \cite[pp.~669 and 673, for
  instance]{Einstein49a} and are, therefore,  not necessarily understood
or shared in the same way by all the different schools. Even more, and
not only in epistemology but in the whole philosophy as well, it is
quite common to find arguments for or against realism that presuppose
the ``existence'' of what they are supposed to accept or deny.  Indeed, what
sense can be given to an affirmation like ``reality does (not)
exist'' ?  In less caricatured sentences, like ``an object is
independent of the subject'' or ``objective facts are an illusion'',
the meaning of ``object'', ``subject'', ``fact'' or even ``illusion''
and the use of the verb ``to be'' reflect some kind of reality
(otherwise we fall in a vicious circle, paved with liar-like paradoxes
 where illusion is itself an
illusion, etc. This argument is also approved 
by Fine, \citeyear{Fine96a}, 
chap.~7, \S~3, first paragraph and discussed by Nozick, \citeyear{Nozick01a}, chap.~1).}

\am{Aside from practical (``fitted'' as Darwin would have said,
``convenient'' as Poincar\'e should have preferred) everyday life, even
an ultra anti-realist like a solipsist must accept the conclusion of
the famous Descartes'credo ``I think, therefore I am''\footnote{I
  leave as an exercise for the reader to explain, from the hints given
  throughout the present article, why it is preferable to invert the
  implication and say ``I am, therefore I think''.}. Nominalists must
accept the existence of equivalence classes that each word
denotes. Agnostic empiricists, phenomenologists, positivists,
pragmatists, idealists, etc.  must deal with the existence of facts,
sensations, mental representations and so on.  Post-modernists of
every kind use equivalence classes like historical
events, social structures and groups, cultural beliefs, etc.,  and 
whether contingent or not, these classes may nevertheless
 become true objects of scientific
studies as soon as they contribute to a rational scheme (the social or
human sciences).  As far as rationality is concerned---and, I would
say, by definition of what rationality means---the best we can do is
to remain at least non self-contradictory, in a \textit{virtuous
  circle} so to speak. My preference in this matters inclines to
distribute the notion of existence into a sort of holistic web where
entities have a more or less high \emph{degree of reality} according
to how tightly bounded they are by coherent and logical relations.  I
will try to show how the four facets of symmetry, through the cardinal
concept of equivalence classes, help us to gain in coherence or, at
least, help us to anchor and place in this circle.  Anyway, wherever
we stand, an Ouroboros loop remains unavoidable because we, as
intelligent subjects, are fully part of the world and  may be, therefore,
considered as objects of knowledge as well. }

\am{The two oil paintings by
Ren\'e Magritte entitled ``La condition humaine'' (the human
condition) could have been chosen to illustrate the complex relation
between real and its representation, but its circularity is even more
admirably rendered by Maurits Cornelis Escher's lithography
``Prentententoonstelling'' (print gallery, see  \cite{deSmit/Lenstra03a}.}

\am{Of course, I can take on safely all the self-reference aspects of the
present reflections since, as being hopefully rational, they are
therefore partly recursive (rational thinking must obviously be broad
enough in scope to embrace the process of rational thinking itself):
to play with the multiple senses of the word ``reflection'' and to use
a metaphor borrowed from laser physics, I hope that, through this
paper, some light will be produced or at least coherently amplified,
like the beam in a resonant optical cavity.}
 This somehow risky and ambitious
 attempt
to enlighten different other fields by extrapolating some
notions whose most precise meaning can only be offered by 
 mathematics and physics  was already a
strategy proposed by Helmholtz in 1868, according to
\citet{Cassirer44a}; the latter analyses carefully the precautions that
must be taken to achieve this goal precisely with the help of the
notion of symmetry. 

\section{Symmetries, classifications and hierarchies}

The physical meanings of the notion of symmetry will be our guidelines
throughout this essay. However, I will start by a rather formal
considerations where the physical motivations remain in the
background.  It is only from the next sections that I will develop the
physical interpretations and extend to a broader domain the matter
of the present section.  After the first version of this work was
composed I became aware of \citeauthor{Brading/Castellani03a}'s
book.  The deep connection between symmetries, classifications and
groups that will be shown in~\S~\ref{subsec:classification} is already
present in Gordon \citet[\S~5]{Belot03a} and Elena
\citet[\S~2]{Castellani03a}'s contributions. Some considerations on the
three first facets have  already been proposed  
by \citet[\S~X.3]{vanFraassen89a}.

\subsection{The four facets of a symmetry}\label{subsec:fourfacets}

The oldest known trace of symmetry involved in a thought process are
the engraved pieces of red ochre found at Blombos Cave in South Africa
and are estimated to be older than 75,000 years
\citep[fig.~2]{Henshilwood+02a}.  Most likely, we will never know
whether these engravings had a symbolic role or were a pure
aesthetical game. However, historically, the concept of symmetry was
first employed to refer to an aesthetic harmony made of unity, rhythm
or balanced proportions emanating from internal or external relations
\cite{Weyl52a,Tarasov86a,Mouchet13a}, for instance in a sense
exquisitely expressed by Charles Baudelaire in his famous poem
\textit{Correspondences}.

In a scientific context these associations are formulated with some
mappings~$\ftransfo{T}$, defined on a set~$\mathscr{E}$ of elements. A
\textit{symmetry} associated with the set of mappings $\mathcal{T}$
encapsulates four significations that are more or less clearly
distinguished in the literature:

\renewcommand{\theenumi}{(\kern -0.15ex{\roman{enumi}})}
\begin{enumerate}
\item A \textit{transformation} $x\mapsto \ftransfo{T}(x)$ where we
  retain the image~$\T{x}\DEF\ftransfo{T}(x)$ of an element (or a set
  of elements)~$x$ by one mapping~$T$ belonging to~$\mathcal{T}$.

\item A \textit{comprehension}: We retain all the images of~$x$
obtained when applying all the mappings in~$\mathcal{T}$ and collect
them in the set
\begin{equation}\label{def:class}
  \sigma(x)\DEF\{\ftransfo{T}(x)\}_{\ftransfo{T}\in\mathcal{T}}\;.
\end{equation}
When applied to many (and ideally all) elements of~$\mathscr{E}$ 
 we can see  the mapping~$x\mapsto\sigma(x)$ as being a 
\textit{classification}. 

\item An \textit{invariance}: Provided~$\mathcal{T}$ fulfills some conditions that will
be specified below, applying any
mapping~$\ftransfo{T}$ of~$\mathcal{T}$ on any subset 
of~$\sigma$ gives again a subset 
of~$\sigma$. In the case of a mirror symmetry for instance, 
the action of~$\ftransfo{T}$ on the 
pair~$\sigma(A)=\{A, \T{A}\}$ where~$A$ is a set of points
of space does not modify $\sigma(A)$.

\item A \textit{projection}\footnote{Mathematicians would 
certainly prefer the term ``section'', keeping the term projection 
to  exclusively qualify 
a mapping equal to its iterates. They would also talk of ``orbit''
rather than of ``comprehension''.} or \textit{symmetry breaking}: 
Rather than dealing with the invariant  sets 
(also called globally invariant or symmetric sets)
$\sigma_1$, $\sigma_2,\dots$
we retain only one element for each $\sigma$ i.e.
$x_{1}, \,x_{2},\dots$ such that~$x_1\in\sigma_1, x_2\in\sigma_2,\dots $  Formally, we can see the projection 
as a mapping~$\sigma \mapsto x$ where an $x$ is chosen 
to be a representative 
of each~$\sigma$.
\end{enumerate}

Projection constitutes the inverse operation of comprehension and
rather than the widely acknowledged expression \textit{symmetry
  breaking}, it should have been more precise to talk about
\textit{invariance breaking}.

\subsection{Symmetry as a classification tool and vice versa}\label{subsec:classification}

Recall that any classification requires the fundamental mathematical
concept of \textit{equivalence relation}: it is a binary relation,
hereafter denoted by $\equiv$, between elements of a set~$\mathscr{E}$
which is, by definition \citep[\S~II.6.1 p.~113 to quote one of the
  paragons of structuralist mathematical books]{Bourbaki68a},
\begin{description}%\setlength{\itemsep}{-.25\baselineskip}   
\item[a)~reflexive\!\!]: every $x$ in $\mathscr{E}$  is related to itself,  $x\equiv x$;
\item[b)~symmetric\!\!]\footnote{It is not a coincidence if we recover this term here!}: for all $(x,y)$ in $\mathscr{E}^2$, if $x\equiv y$ then $y\equiv x$;
\item[c)~transitive] (Euclide's first common notion): for all $(x,y,z)$ in $\mathscr{E}^3$ if $x\equiv y$  and $y\equiv z$ then $x\equiv z$.
\end{description}
From a set of mappings~$\mathcal{T}$ acting on~$\mathscr{E}$  we can define
\begin{equation}\label{def:equiv}
  \parbox{7cm}{$x$ to be related to~$y$
if and only if there exists a mapping~$\ftransfo{T}$
in $\mathcal{T}$  such that $y=T(x)$.}
\end{equation}
Saying that this relation is
 reflexive means that there exists a transformation 
in~$\mathcal{T}$ for which $x$ is invariant. A sufficient condition
is of course that

\noindent a')~$\mathcal{T}$ includes the identity mapping.

\noindent The symmetry property is fulfilled if  

\noindent b')~Every
mapping is invertible and its inverse must belong to $\mathcal{T}$.

\noindent Finally, a sufficient condition to have the 
transitivity property of the binary 
relation is the closure property
for~$\mathcal{T}$:

\noindent c')~The composition
$T_1\circ T_2$ of two mappings belonging to
$\mathcal{T}$ remains in $\mathcal{T}$.

\noindent Properties a'), b'), c') are rather intuitive when 
using the language of transformations: 

\noindent a')~Doing nothing is a (somehow trivial) transformation; 

\noindent b')~Every transformation 
can be undone; 

\noindent c')~Making two successive transformations is a transformation. 

\noindent Together with the associativity 
of the composition law---we always have
$(T_1\circ T_2)\circ T_3=T_1\circ (T_2\circ T_3)$--- they define 
the set~$\mathcal{T}$ as being a \textit{group}.
The set~$\sigma(x)$ defined by~\eqref{def:class} is called the 
\textit{equivalence class} of~$x$ and
 collects all the elements that are equivalent to $x$ i.e. related 
by the equivalence relation. 

Summing up, we have seen how symmetry and classification are
intimately related and why the group structure arises naturally .

\textbf{Remark:} It is conceptually interesting to remark that not
only working with a group of transformations is sufficient for having
a classification, but it is also a necessary condition in the sense
that, given a classification, we can always construct an ad hoc group
of transformations~$\mathcal{T}$ such that~\eqref{def:equiv} for the a
priori given equivalence relation
 \citep[see also][\S~X.3]{vanFraassen89a}.  For example we can
choose~$\mathcal{T}$ as the group generated by all the one-to-one mappings
(bijections) that are the identity everywhere but on one class. There
is also the mathematical possibility of working with a
set~$\mathcal{T}$ that is not a group but for which~\eqref{def:equiv}
is still an equivalence relation. For instance take for~$\mathcal{T}$
the set of all mappings~$\ftransfo{T}_\sigma$ whose restriction to one
class~$\sigma$ is a bijection and that sends any element belonging to $\sigma'\neq\sigma$ to one
chosen representative element of~$\sigma'$. Except for degenerate
cases where there is only one class or when all but one of them have
exactly one element, $\mathcal{T}$ does not in general possess the closure
property, does not contain the identity and the~$\ftransfo{T}_\sigma$
are not invertible\footnote{I am grateful to Emmanuel Lesigne for
  helping me in clarifying these points.}. In physics I do not know
any relevant examples of such a situation.

\subsection{Hierarchies}\label{subsec:hierarchies}

The procedure of constructing new mathematical \textit{objects} as
being equivalence classes is omnipresent in mathematics.  The set of
equivalence classes in~$\mathscr{E}$ constructed from the
relation~$\equiv$ constitutes the so-called quotient set denoted
by~$\mathscr{E}/\equiv$.  Just to give one fundamental example: a
hierarchy of numbers can precisely be built with equivalence classes
defined at each step with an appropriate equivalence relation: the
rational numbers being equivalence classes of ordered pairs of
integers\footnote{By the way, in the Frege construction, even natural
  numbers are equivalence classes, namely the equivalence classes of
  sets related by one-to-one mappings \citep[\S\S~III.3.1 p.~157 and
    III.4.1 p.~166]{Bourbaki68a}.  As far as physics is concerned, I
  will always stay at the level of ``naive'' set theory without taking
  the subtle precautions which avoid paradoxes \`a la Russell, in
  particular by distinguishing between the so-called proper classes
  and sets. Perhaps, category theory should provide
 a more suitable
 framework, specially if one is akin to structural realism.  
As far as logical considerations are concerned, I mention
  that I will also take for granted the axiom of choice that allows
  the operation of projection.  }, the real numbers being equivalence
classes of (Cauchy) sequences of rational numbers (and 
appear, therefore, as classes constructed from classes), the unit-modulus numbers
can be seen as classes of real numbers considered as being equivalent
if they differ by an integer multiple of $2\pi$ (classes from classes
from classes).  More generally, we first define a new
set~$\mathscr{E}'_1$ of elements constructed from the elements
of~$\mathscr{E}_1$ and then we classify them
in~$\mathscr{E}_2=\mathscr{E}'_1/\equiv$ using an equivalence
relation~$\equiv$ defined in $\mathscr{E}'_1$.  Another standard
example is given in Euclidian geometry:~$\mathscr{E}_1$ is the set of
points, $\mathscr{E}'_1=\mathscr{E}_1\times\mathscr{E}_1$ is the set
of ordered pairs of points and, eventually, vectors appear as equivalence
classes of ordered pairs of points, two pairs of points being equivalent if
and only if they form a parallelogram.

The objects formed this way gain a sort of autonomy with respect to
the primitive elements from which they are built. The different
operations or relations that may exist in~$\mathscr{E}_1$ can be used
to define operations in~$\mathscr{E}_2$ provided that they do not depend on
the choice of the representative elements in the classes. Therefore,
computations in~$\mathscr{E}_2$ can often (but not always, see the
last paragraphs of \S~\ref{subsec:invbreak}) be done without any
reference to computations at the ``lower'' level
in~$\mathscr{E}_1$. To keep working with the same previous example, in
a $d$-dimensional space, one can add two vectors using their $2\times
d$ coordinates, not using the $2\times2\times d$~numbers that encode
the position of each points of the two representative pairs.  Then,
when constructing the upper level~$\mathscr{E}_2$ the information that
encodes the distinction between all the elements of a class at the
lower level is erased. In other words the ``internal position'' of an
element in a class has been made irrelevant or superfluous
\citep[\S~3]{Ismael/vanFraassen03a,Castellani03a}  as far as the
manipulations of the elements of~$\mathscr{E}_2$ are concerned.

\enlargethispage*{\baselineskip}
Moreover, the main motivation of constructing equivalence classes is
to obtain a richer structure. One can therefore truly speak of
\textit{mathematical emergence} in the sense that we can draw out
properties in~$\mathscr{E}_2$ that are not relevant
for~$\mathscr{E}_1$. The emergent notions (e.g. continuity or Borel
measure) concern the elements of~$\mathscr{E}_2$ (e.g. the real
numbers) rather than the elements of~$\mathscr{E}_1$ (e.g. the integers).
\begin{quotation}\small{According 
to the modern conception advocated by Klein, the 
characteristic properties of a multiplicity must 
not be defined in terms of the \emph{elements} of 
which the multiplicity is composed, but solely in 
terms of the \emph{group} to which the multiplicity is related.[\dots]
The real foundation of mathematical certainty
lies no longer in the elements from which mathematics
 starts but in the \emph{rule} by which the elements are related
to each other and reduced to a ``unity of thought''
}\citep[\S~II, pp.~7--8]{Cassirer44a}.
\end{quotation} 

\begin{figure*}[!ht]
\includegraphics[width=16cm]{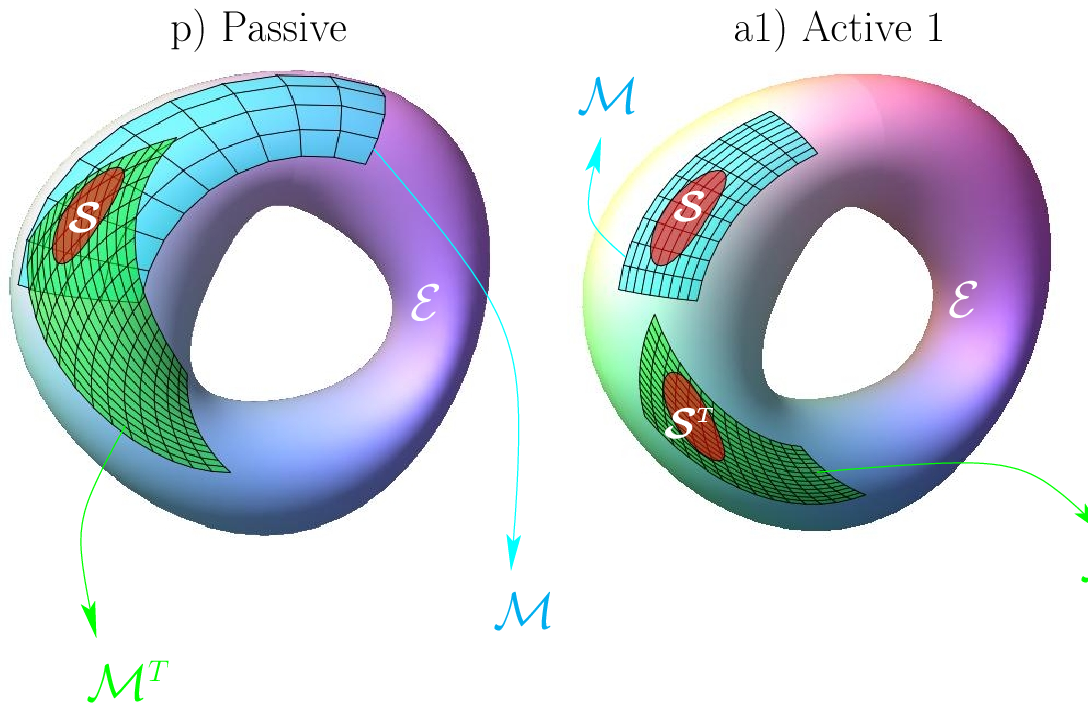}
\caption{\label{fig:transfo_passive_active}Schematic view of the three
  physical interpretations of a transformation: p) In a passive
  transformation the same system~$\mathcal{S}$ is described by two
  different measurers~$\mathcal{M}$ and~$\T{\mathcal{M}}$.  For
  instance, they use two kinds of measuring instruments, different
  gaugings or standards and two different frames of reference to map
  the space-time around~$\mathcal{S}$.  In order to give a sense to
  ``same'' and ``different'' used above, it is necessary to use an
  environment~$\mathcal{E}$ as a reference.  In the active points of
  view~$\mathcal{S}$ is turned into~$\T{\mathcal{S}}$ and
  either~$\mathcal{M}$ is transformed accordingly (a1) or remains
  unchanged (a2).}
\end{figure*}
\section{Transformations}\label{sec:transformations}

\subsection{Active and passive physical interpretations of a transformation}
\label{subsec:activepassive}

Let us now sew some physical flesh onto the mathematical bones I have
introduced in the previous section.  Let us begin with the first
facet. As far as I know the distinction between passive and active
interpretations of a transformation was made explicitly for the first
time by \citet*[\S~2.4a]{Houtappel+65a}.  Furthermore I will
follow \citet[\S~1.7]{Fonda/Ghirardi70a} and will consider that
two active points of view can be distinguished.  I will emphasize that
the three physical interpretations of a transformation require the
division of the universe in three parts, not two: the system~$\mathcal{S}$
(the object), the measurer~$\mathcal{M}$ (the subject, the observer)
who describes quantitatively~$\mathcal{S}$ with the help of measuring
devices, an environment~$\mathcal{E}$ made of everything else and that
is necessary to detect whether effectively~$\mathcal{S}$ or~$\mathcal{M}$
have been transformed or not (figure
\ref{fig:transfo_passive_active}).  In a passive
transformation,~$\mathcal{S}$ remains anchored to~$\mathcal{E}$ and
the measurer is transformed to~$\T{\mathcal{M}}$ (for instance the
measuring instruments, the reference frame, the standards are
changed).  In the active points of view the system is changed with
respect to~$\mathcal{E}$ (in cosmology since, in principle, we
consider~$\mathcal{S}=\mathcal{E}$, the active points of view are
meaningless). Unlike in the first active point of view where
both~$\mathcal{M}$ and $\mathcal{S}$ are changed, in the second active
point, the observer~$\mathcal{M}$ remains unchanged. In the latter,

\begin{quote}\small{
[The observations are] of the same 
observer, using one definite language
to characterize his observations} \citep[\S~2.4a, p.~601]{Houtappel+65a}.
\end{quote}

One can find an echo of such different points of view at the
mathematical level.  For instance, in differential geometry, (active)
transformations, like parallel transport, of geometrical objects such
as vectors or forms defined on a manifold are conceptually different
from (passive) coordinates changes.  However, in physics (and in all
empirical sciences), active and passive transformations have much
stronger conceptual differences (\S\S~\ref{subsec:movingboundaries}
and \ref{subsec:reprod_object}) as well as important operational
differences (\S\S~\ref{subsubsec:opdiff} and \ref{subsec:poincare}).

\subsection{Moving the boundaries}\label{subsec:movingboundaries}

Since~$\mathcal{S}$ and~$\mathcal{M}$ necessarily interact, the
threshold between them can be a matter of convention: in the
Michelson-Morley experiment, we may decide freely if the mirrors of
the interferometer are included in~$\mathcal{S}$ or in $\mathcal{M}$.
As noted by \citet[\S~VI.1]{vonNeumann55a}, this convention is
linked to the choice of what facts are considered to be observed,
registered and acknowledged by~$\mathcal{M}$.  For instance during an
observation of the Sun, $\mathcal{M}$ can consider that he counts the
sunspots directly or, rather, that he counts the dark spots that
appear in the projection screen he uses; he can measure the degree of
excitation of the retina cells of another human being looking through a
telescope or look on a functional Magnetic Resonance Imaging
 scanner screen the activity of the
corresponding brain areas during the observation.  Solving the a
priori ambiguity of what a ``direct'' observable phenomenon or a
``sense datum'' is consists in choosing where we place, in a whole
chain of processes, the border between~$\mathcal{S}$ and~$\mathcal{M}$
that defines the so-called ``external world'' (implicitly external
to~$\mathcal{M}$, for a comment on this expression see
Sankey, \citeyear[note~3]{Sankey01a}).  Consequently, before the
interpretation in terms of \textit{correlations} between data and in
terms of concepts forged to cement them, this operation of cutting the
chain constitutes an unavoidable task of science whose difficulty
measures how remote we stand from our human sensations and from our
original evolutionary-shaped abilities (the so-called intuition or
common sense).
\begin{quotation}\small{The important consequence is that, so far,
we are left without criteria which would enable us to draw a
non-arbitrary line between ``observation'' and ``theory''. Certainly,
we will often find it convenient to draw such a
to-some-extent-arbitrary line; but its position will vary widely from
context to context [\dots] But what ontological ice does a mere
methodologically convenient observational-theoretical dichotomy cut?

Does an entity attain physical thinghood and/or ``real existence'' in
one context only to lose it in another?  }
  \citep[pp.~7--8]{Maxwell62a}.\nocite{Feigl/Maxwell62a}\nocite{Savage90a}
\end{quotation} 
In section~\S~\ref{subsec:stabilityabstraction} below, I will propose an 
attempt, based on equivalence classes, to clarify this ontological point.

Besides, quantum physics renders these issues more problematic: since
no generally well-accepted quantum theory has been proposed where the
unicity of the result of \textit{one} measurement emerges from purely
quantum processes, we still do not know how exactly the transition
between quantum systems and classical measuring apparatus occurs.  In
the absence of a more satisfactory solution, we are forced to keep the
orthodox approach where Schr\"odinger equation is replaced by a
discontinuous transition at the very moment of
measurement\footnote{\label{fn:measurement}In the last decades,
  significant progress have been made both theoretically and
  experimentally which focus on the role played by the environment
  during quantum evolution. Among the ocean of bibliographic
  references on this subject, let us mention
  \citet{Wheeler/Zurek83a,Presilla+96a,Breuer/Petruccione02a,
    Giulini03+a, Schlosshauer07a, Barchielli/Gregoratti09a} and their
  references. However, I think the central problem of quantum
  measurement discussed in \S~\ref{subsec:problems} still remains.}.
In what follows, since I will not specially be concerned by these
quantum measurement issues (except in \S~\ref{subsec:problems}), I
will not distinguish between the apparatus that prepares the system
and the detectors.  Within the very general level of the present
discussion, it is not important to specify which part of the degrees
of freedom belongs to~$\mathcal{S}$ or to $\mathcal{M}$ and we will
not be concerned by their entanglement if any (the
environment~$\mathcal{E}$ may have disentangled them according to the
decoherence process).  Furthermore, as noticed by Russell:
\begin{quotation}\small{There has been a great deal 
of speculation in traditional philosophy which might have been avoided
if the importance of structure, and the difficulty of getting behind
it, had been realised. For example, it is often said that space and
time are subjective, but they have objective counterparts; or that
phenomena are subjective, but are caused by things in themselves,
which must have differences \emph{inter se} corresponding with the
differences in the phenomena to which they give rise. Where such
hypotheses are made, it is generally supposed that we can know very
little about the objective counterparts. In actual fact, however, if
the hypotheses as stated were correct, the objective counterparts
would form a world having the same structure as the phenomenal world,
and allowing us to infer from phenomena the truth of all propositions
that can be stated in abstract terms and are known to be true of
phenomena.[\dots] In short, every proposition having a communicable
significance must be true of both worlds or of neither.} \citep[chap.~VI, p.~61]{Russell19a}.
\end{quotation} 
I will therefore not consider the separation between~$\mathcal{S}$
and $\mathcal{M}$ to be of metaphysical origin and I will not endorse
any kind of genuine dualism nor pluralism but prefer a realistic
(materialistic) monism: The separation between $\mathcal{S}$ and
$\mathcal{E}$ (that demarcates the pertinent parameters from the
irrelevant ones) together with the separation between $\mathcal{M}$
and $\mathcal{S}$ consist not only of blurred but above all of movable
boundaries that must be moved to check the consistency of our line of
reasoning (for an interesting proposal that trends to reinforce the
conception of a unified ontology, see Clark, \citeyear{Clark08a}). For
instance, when the effects of the coupling to the environment are
under the scope, this may be implemented by splitting~$\mathcal{S}$ in
two: a subsystem coupled to a bath that furnishes a reduced model of
the ``exterior''.
\enlargethispage*{\baselineskip}
\begin{quotation}\small{Where does that crude reality, 
in which the experimentalist lives, end, and where 
does the atomistic world, in which the idea of reality is illusion and
anathema begin ? There is, of course, no such border; if we are compelled
to attribute reality to the ordinary things of everyday life including scientific instruments and materials used in experimenting, we cannot 
cease doing so for objects observable only with the help of instruments.
To call these subjects real and part of the external world does not, 
however, commit us in any way to any definite description: 
a thing may be real though very different from other things we know.
[\dots]\\
\indent The boundary between the action of the subject 
and the reaction of the object is blurred indeed. 
But this does not prohibit us from using these concepts in a reasonable way.
The boundary of a liquid and its vapour is also not sharp, as their 
atoms are permanently evaporating and condensing. 
Still we can speak of liquid and vapour\footnote{No doubt that
Born was also aware that one can bypass the coexistence curve and
circumvent the critical point 
in the pressure-temperature diagram by
 \textit{continuously} connecting the liquid phase  with
vapour while remaining at the macroscopic scale all the way around. }
}\citep{Born53a}.
\end{quotation}
Indeed, since Samuel Johnson's refutation of Berkeley's extreme
idealism---obtained by merely ``striking his foot with mighty force
against a large stone, till he rebounded from it''\citep[6 august
  1763]{Boswell1791a}---, the possibility of action on the world has
always been an insight of what reality means. Active and passive
transformation just help to extend, precise and formalise this common
sense (by the way, the use of the word ``common'' in this expression
already bears the notion of invariance).

\subsection{Reproducibility and objectivity}\label{subsec:reprod_object}

On the one hand, active points of view are at work when one wants to
put to the test the \textit{reproducibility} of an experiment.
Galilei's ship (\citeyear{Galilee1632b}, 2nd day, 317)---see also its
inspiring predecessors like Bruno's ship thought
experiment (\citeyear{Bruno1584a}, third dialogue) borrowed from an older
argument used by %Nicholas of 
\citet[II.12, p.~111]{Cusa1440}---is probably the most famous 
and, in a historical
  perspective, certainly the most crucial illustration of a (first)
  active transformation.  An example, among many others, of the second
  active point of view can be found in Gibbs'interpretation of a
  thermodynamical ensemble where we consider ``a great number of
  independent systems'' \citep[chap.~I, p.~5 and the
    preface]{Gibbs10a}. As usual in classical pre-twentieth century
  tradition, the reference to any subject~$\mathcal{M}$ is implicit
  because it is considered as being irrelevant (one noticeable
exception being discussions on the Maxwell's demon).
  
Of course, from one experiment to another, time may have evolved. This
is reflected in (or equivalent to) many transformations that
occurred in the environment~$\mathcal{E}$.
Whether the experiment is actually reproducible or not 
will be a hint on the relevance of the part of the universe
we have decided to consider as the system~$\mathcal{S}$.

On the other hand, the passive point of view puts to the test the
\textit{intersubjectivity} of an observation and therefore paves the
road to \textit{objectivity} (see Nozick's \citeyear{Nozick01a}, chap.~2, p.~91).
 An invariance under a passive
transformation reveals or defines an entity as being (at least
approximately) independent of~$\mathcal{M}$.  To which extent this
autonomy appears, and more generally the degree of reality
of~$\mathcal{S}$, is reflected in the nature of the group of
transformations~$\mathcal{T}$ for which invariance occurs.

Before I specify~$\mathcal{T}$ in the next subsection, from what
precedes we can already see how a bridge can be established between
two main principles of the scientific method, reproducibility and
objectivity: for this purpose, symmetry is the appropriate tool (we
shall come back on this point below).

\subsection{Relevant groups of transformations}\label{subsec:relevantgroups}

Up to now, the strong affinity between classification and
transformations which was shown in \S~\ref{subsec:classification} may
appear of shallow interest to the physicist. Actually, one can
classify anything with anything or even transform anything to anything
else.  When a property~$P$ is given a priori ``Having the same
property~$P$ as'' defines automatically an equivalence relation and
the corresponding equivalence classes for which~$P$ is an invariant;
reversely, if a partition of a set is given, the property~$P$ can
always be identified with the membership to a subset and the
equivalence relation defined as ``belongs to the same subset as''.
The transformations that are most pertinent are those that can be
applied to a wide range of entities while keeping simple rules, that
is, defined with less information than the set of elements they act on
(we will come back in \S~\ref{subsubsec:pruning} 
to the reduction of information provided by
any classification and therefore any symmetry).  A reversible
transformation that turns a pumpkin into a coach, whether it concerns
the three-dimensional objects or a morphing between two-dimensional
images, must encapsulate the huge amount of information than encodes
the pumpkin and the coach; it can hardly be applied to anything else
without artificially introducing some additional ad hoc
information. This is why, scientifically speaking, such a
transformation is far less interesting than a rotation or a
dilatation.

\subsubsection{Algebraic representations}\label{subsubsec:algrep}

\begin{figure}[!ht]
\begin{center}
\includegraphics[width=7cm]{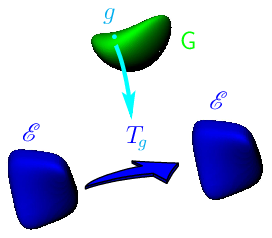}
\caption{\label{fig:representation}Schematic view of
the algebraic representation of an abstract group~$\mathsf{G}$ in a set
of elements~$\mathscr{E}$. To each element~$g$ in~$\mathsf{G}$
is associated a transformation~$T_g$ acting on the elements of~$\mathscr{E}$.}
\end{center}
\end{figure}

Indeed, the relevant groups of transformations that are considered in
physics have been abstracted independently of the elements that are
transformed.  For instance, the usual transformations like
space-translations or rotations have their roots in our intuitive
geometrical conception of the three-dimensional space; yet, in
mathematical physics, they have now acquired the status of an
``abstract'' group~$\mathsf{SO}(3)$ that can be \textit{represented}
in many ways: they do not apply to three-dimensional geometrical
objects only but also to a huge collection of mathematical objects
living in much more abstract spaces (specially the quantum states and
operators). Figure~\ref{fig:representation} provides a schematic view
of what an algebraic representation of an abstract group is.  The
theory of abstract group and their representations constitutes a whole
domain of algebra and mathematical physics (among many treatises, see
for instance Cornwell, \citeyear{Cornwell84a}; Sternberg,
\citeyear{Sternberg94a} or Jones, \citeyear{Jones98a}. The special
status of linear representations in quantum physics can be traced back
to a theorem due to Wigner in the early 30's, see the references of
Simon et al, \citeyear{Simon+08a} and Mouchet \citeyear{Mouchet13b}.).

Many abstract symmetry groups have been shown to be relevant for the
six purposes $1^\circ$--$6^\circ$ listed in
\S~\ref{sec:introduction}. Let us direct our attention to several
substantial extensions that go far beyond the set of pure spatial
isometries:

\paragraph{Evolution:} First, the time evolution must be 
considered as a transformation\footnote{Technically, if 
we are to describe irreversible 
processes
we can relax b') in \S~\ref{subsec:classification}
 and work with a so-called semigroup.}.  This strengthens the interpretation
of the transformation in terms of reproducibility since the latter
concerns space-translations as well as time-translations. This is also
in accordance with Einsteinian relativity that treats on an almost
equal footing time and space.

\paragraph{Transformations mixing space and time:} Second, in accordance with
Galileo's ship argument, we can also include dynamical transformations
corresponding to translations at constant speed (the transformation of
spatial positions depends on time).  General relativity incorporates
any smooth transformation mixing space and time. Canonical formalism
allows for even more general transformations involving mixture
of spatio-temporal coordinates and momenta.

\paragraph{Exchange of particles:}
As suggested first by \citet[in particular
  eq.~8]{Heisenberg26a,Heisenberg27a} and \citet{Dirac26a}, the
permutations (or ``substitutions'') of a finite number of elements
(for which the term of ``group'' was originally coined by Galois in
1830) appeared of crucial importance for describing quantum collective
effects of indistinguishable particles. Later on,
\citet{Heisenberg32a} considered the exchange of non-identical
particles, namely protons and neutrons, to describe the spectral
properties of nuclei.  Another important example of transformations
involving changes of particles goes back to Dirac's \citeyear{Dirac30a}
seminal work that led to the charge conjugation symmetry
\citep[for instance, see][]{Yang94a} where particles may be
substituted with their antiparticles.  Even though there is still no
experimental evidence of its relevance, since the 1970's, the
transmuting of bosons into fermions and conversely have been
introduced in several particle and string models.  This concept of
supersymmetry is an attempt to tackle some problems raised by the
Standard Model \citep[\S~24.2]{Weinberg00a},
\citep[chap.~1]{Binetruy06a}.

\paragraph{Gauge transformations:} Field theories that describe
local interactions in condensed matter or in high energy physics, may
require the introduction of some mathematical ingredients whose
choice, in a finite portion of space-time, is, to a large extent, a
matter of convention. A gauge transformation represents a change of
convention and is, by definition of what a convention means, not
expected to have any observable influence.  Looking
systematically for the effects of these transformations on the
mathematical ingredients of a model provides sufficiently constrained
guidelines to select the physically relevant descriptions.  There is
no need to insist any longer on the efficiency of such constructions
throughout the twentieth century, from the elaboration of general
relativity and Weyl's first attempts starting in 1918, to
\citet{Yang/Mills54a} non-commutative gauge model and
beyond. Among the vast literature on the subject see, for instance,
\citet{Moriyasu83a} for a nice introduction.

\subsubsection{Operational differences between active and 
passive transformations}\label{subsubsec:opdiff}

All the transformations described above have been upgraded into
abstract groups and can be algebraically represented within the
different species of the mathematical ingredients of various physical
models.  However, if some transformations can be indifferently
interpreted in both passive and active point of view, this is far from
being the case for all of them.

Precisely because gauge transformations fulfill their constructive
role when they have no detectable consequences, they can only be
interpreted as passive transformations \citep[and its
  references]{Brading/Brown04a}. The same situation occurs in
classical mechanics: after a relevant canonical transformation that
interweaves them, the dynamical variables may loose the empirical
interpretation they had before being transformed (say, the positions
and velocities of the planets when blended into action-angle
variables).  Therefore, such transformations appear clearly as a change
of description of the dynamics rather than an active transformation of the
system.

Examples of active (resp. passive) transformations that cannot
naturally be considered from the passive (resp. active) point of view
are given by transformations associated with an exchange, addition or
subtraction of some constituents of the system (resp. the measuring
device).  If we are ready to accept a permutation between quantum
particles as a candidate for being in~$\mathcal{T}$, why should we not
consider also the replacement of a spring in a mechanical apparatus or
the modification of the wavelength of a probe?  Even though, these
kinds of transformations are never (as far as I know) considered when
talking about symmetry, this is precisely invariance with respect to
them, as being discussed in \S~\ref{sec:invariance} below, that allows
eventually to establish stable and universal laws.  In practise, when
attempting to reproduce an old or a remote experiment, the
experimental set up is rarely moved but rather completely rebuilt from
new materials.

\subsection{Perception of space-time transformations: following Henri 
Poincar\'e's intuitions}\label{subsec:poincare}

In fact, a way of understanding which mathematical transformations can
be interpreted both passively and actively was presented more a
century ago by Poincar\'e (\citeyear[\S~5]{Poincare1895a,Poincare03a})
when analysing carefully our intuition of three-dimensional space (his
considerations can be applied to a large extent to the four-dimensional space-time).  He introduced a distinction that Einstein
would sum up four decades after him:
\begin{quotation}\small{
Poincar\'e has justly emphasized the fact that we distinguish two
kinds of alterations of the bodily object, ``changes of state" and
``changes of position."  The latter, he remarked, are alterations
which we can reverse by arbitrary motions of our bodies} 
\citep[\S~2, pp.~354--355]{Einstein36a}.
\end{quotation}
However, what does not appear clearly in this
report (but was criticised by Einstein elsewhere, for instance in his 
 \citeyear[p.~677]{Einstein49a} reply)  is that, according to Poincar\'e, this is a matter of 
definition or convention: 
\begin{quotation}\small{
Thence comes a new distinction among external changes: 
those which may be so corrected \underline{we call} changes of position; 
and the others, changes of state}
 \citep[\S~5, I underline]{Poincare03a}.
\end{quotation}
 
\noindent For Poincar\'e, the possible compensation of an active 
transformation by a passive one is therefore intimately linked with 
our perception of space. 

\begin{quotation}\small{
For a being completely immovable there would be neither space nor
geometry; in vain would exterior objects be displaced about him, the
variations which these displacements would make in his impressions
would not be attributed by this being to changes in position, but to
simple changes of state; this being would have no means of
distinguishing these two sorts of changes, and this distinction,
fundamental for us, would have no meaning for him}
  \citep[\S~5]{Poincare03a}.
\end{quotation}

However, following Galileo's conception of uniform speed, Poincar\'e
was aware that transformations could not be considered absolutely and,
as soon as~1902, he clearly formulated in chap.~V, \S~V, of
\textit{Science and Hypothesis} a ``relativity law''.  The
transformations must be anchored somewhere: this is precisely the main
reason why we must introduce the environment~$\mathcal{E}$ as a
background and that \citet[\S~2, p.~354]{Einstein36a} would call
a ``bodily object''.  Otherwise, the first active point of view would
be undetectable and the distinction between the passive and the
second active point of view would be meaningless.  Extending an old
argument going back to Laplace \citep[according to][chap.~5,
  pp.~168--169]{Jammer94a}, Poincar\'e not only
discusses the case of a uniform expansion of the universe but also
considers any continuous change of coordinates:
\begin{quotation}\small{But more; worlds will be indistinguishable 
not only if they are equal or similar, that is, if we can pass from 
one to the other by changing the axes of
 coordinates, or by changing the scale to which 
lengths are referred; but they will still be indistinguishable if 
we can pass from one to the other by any `point transformation' whatever.
[\dots]
The relativity of space is not ordinarily understood 
in so broad a sense; it is thus, however, that it would be proper to understand it}
 \citep[\S~1]{Poincare03a}.
\end{quotation}

\noindent Poincar\'e considers that only \textit{topological}
structures---being unaffected by homeo\-mor\-phisms---are physically
relevant. For him the choice of \textit{differential} structure, in
particular through the metric, is a matter of convention similar to a
choice of units or a choice of language. The imprint of curvature in
the sum of the angles of a triangle or on the ratio between the
perimeter of a circle and its radius can always be rectified
empirically.
\begin{quotation}\small{ Think of a material circle, measure its radius and
  circumference, and see if the ratio of the two lengths is equal
  to~$\pi$ . What have we done? We have made an experiment on the
  properties of the matter with which this \emph{roundness} has been
  realized, and of which the measure we used is made}
\citep[chap.~V, \S~2]{Poincare52a}.
\end{quotation}
 
\noindent and also

\begin{quotation}\small{ The same question may also be asked in another way. If
  Lobatschewsky's geometry is true, the parallax of a very distant
  star will be finite. If Riemann's is true, it will be
  negative. These are the results which seem within the reach of
  experiment, and it is hoped that astronomical observations may
  enable us to decide between the two geometries. But what we call a
  straight line in astronomy is simply the path of a ray of light. If,
  therefore, we were to discover negative parallaxes, or to prove that
  all parallaxes are higher than a certain limit, we should have a
  choice between two conclusions: we could give up Euclidean geometry,
  or modify the laws of optics, and suppose that light is not
  rigorously propagated in a straight line. It is needless to add that
  every one would look upon this solution as the more
  advantageous. Euclidean geometry, therefore, has nothing to fear
  from fresh experiments} \citep{Poincare1891a}.
\end{quotation}

Albeit historical development of the theory of gravitation has refuted
Poincar\'e's opinion
 that
\begin{quotation}\small{Euclidean geometry is, and will remain, the most convenient:
  1st, because it is the simplest, and it is not so only because of
  our mental habits or because of the kind of direct intuition that we
  have of Euclidean space; it is the simplest in itself, just as a
  polynomial of the first degree is simpler than a polynomial of the
  second degree; 2nd, because it sufficiently agrees with the
  properties of natural solids, those bodies which we can compare and
  measure by means of our senses} \citep{Poincare1891a}.
\end{quotation}

\noindent when gravitation was formulated as a gauge theory by
\citet{Utiyama56a}, then working either with Euclidian/Minkowskian
or Riemannian geometry appeared indeed to be a conventional choice. See
\citet[chaps.~15--17]{Carnap66a}, \citet{Stump91a} or
\citet{Hacyan09a} for a more detailed discussion on these
issues. (Even if we put aside the topological properties of space-time, like
worm-holes at large scales or any kind of topological defect that may
have physical observable consequences which would definitely rule out
Minkowskian geometry, some general arguments on causality
still favour a curved space-time interpretation, \citet{Penrose80a}).

Even if we keep sticking to Riemannian geometry, we can remark that a
confusion remains that obscures the debate between realists and
conventionalists (or their relativist post-modern successors).  A
representation (here a choice of one coordinate map) is always,
obviously, a matter of convention (it fixes a gauge); however, when
quotienting all the equivalent descriptions obtained one from each
other by a transformation (the change of coordinates), we can abstract
geometrical objects with intrinsic properties (the Riemannian manifold
itself and the tensors defined on it). Formally from a set of
quantities indexed by coordinates labels, say~$v^{\mu}$, whose values
change when changing the coordinates from~$(x^\mu)_\mu$
to~$(\T{{x^\mu}})_\mu$,
\begin{equation}\label{eq:vectorcoordinates}
  \T{{v^{\mu}}}=\frac{\partial \T{{x^{\mu}}}}{\partial x^\nu } v^{\nu}\,\;,
\end{equation}
an abstract object can be defined, the tangent vector~$v$, which is
independent of the precise choice of coordinates.  While the
coordinates~\eqref{eq:vectorcoordinates} are said to be co-variant
under the (passive) transformation that changes the coordinates, the
vector~$v$ remains insensitive to it. Besides, to emphasize the
intrinsic character of the construction, rather than starting from the
transformation rule~\eqref{eq:vectorcoordinates}, a vector at a
point~$x$ is preferably defined in differential geometry, as being an
equivalence class of tangent differential curves on the manifold
passing through~$x$ \citep[chap,~III, B.1,
  pp.~120--121]{ChoquetBruhat/DeWittMorette82a}.  This geometrisation
procedure was the core of Klein's Erlanger
program (\citeyear{Klein1893a}, see Cassirer, \citeyear{Cassirer44a}, 
for a historical account
  on the connections between a group of geometrical transformations,
  perception, invariance and objectivity. See also the excerpts given
  above in \S~\ref{subsec:hierarchies} and below in
  \S~\ref{sec:invariance}) and one of the fathers of
the modern notion of symmetry was already aware of its physical
connection with objectivity \citep[for a recent
  survey, see][\S~2]{Kosso03a}:
\begin{quotation}\small{Without claiming to give a mechanically applicable criterion, our description bears out the essential fact
that objectivity is an issue decidable on the ground of experience
only.[\dots] Whereas the philosophical question of objectivity is not
easy to answer in a clear and definite fashion, we know exactly what
the adequate mathematical concepts are for the formulation of this
idea.[\dots] The pure mathematician will say: Given a group
$\mathsf{G}$ of transformations in a field of symbols, a geometry is
established by agreeing to study, and consider as objective, only such
relations in that field as are invariant under the transformations of
$\mathsf{G}$ } \citep[\S~13, pp.~72 and~77]{Weyl49a}.
\end{quotation}
\begin{quotation}\small{Perhaps the philosophically most relevant feature of modern
science is the emergence of abstract symbolic structures as the hard
core of objectivity behind---as Eddington puts it---the colourful tale
of the subjective storyteller mind} \citep[Appendix~B, \S~1,
    p.~237]{Weyl49a}.
\end{quotation}
and also\enlargethispage*{2\baselineskip}
\begin{quotation}\small{But I preferred the name of
automorphisms [\dots], defining them with Leibniz as those
transformations which leave the structure of space
unchanged.[\dots]\\ \indent We found that objectivity means invariance
with respect to the group of automorphisms.  Reality may not always
give a clear answer to the question what the actual group of
automorphisms is, and for the purpose of some investigations it may be
quite useful to replace it by a wider group. For instance in plane
geometry we may be interested only in such relations [that] are
invariant under parallel or central projections; this is the origin of
affine and projective geometry.  The mathematician will prepare for
all such eventualities by posing the general problem, how for a given
group of transformations to find its invariants (invariant relations,
invariant quantities, etc.), and by solving it for the more important
special groups---whether these groups are known or are not known to be
the groups of automorphisms for certain fields suggested by
nature. This is what Felix Klein called "a geometry" in the abstract
sense. A geometry, Klein said, is defined by a group of
transformations, and investigates everything that is invariant under
the transformations of this given group} \citep[chap.~II, p.~42 and
    chap.~IV, pp.~132--133]{Weyl52a}.
\end{quotation}

Poincar\'e's conventionalism far from being antithetic to objectivity \citep[see][]{Giedymin82a,Giedymin92a}
appears to be a necessary condition for the latter:
\begin{quotation}\small{Now the possibility of  \underline{translation} implies the existence of an
   \underline{invariant}.  To translate is precisely to disengage this
  invariant.[\dots]\\ 
  \indent What is  objective must be common to many
  minds and consequently transmissible from one to the other, and this
  transmission can only come about by [a] ``discourse'' [\dots] we are
  even forced to conclude: no discourse, no  \underline{objectivity}.
  [\dots]\\ 
  \indent Now what is science? I have explained [above], it
  is before all a \underline{classification}, a manner of bringing together facts
  which appearances separate, though they were bound together by some
  natural and hidden kinship. Science, in other words, is a system of
  \underline{relations}.  Now we have just said, it is in the relations alone that
  objectivity must be sought; it would be vain to seek it in beings
  considered as isolated from one another.\\ 
  \indent To say that
  science cannot have objective value since it teaches us only
  relations, this is to reason backwards, since, precisely, it is
  relations alone which can be regarded as objective.\\ 
  \indent External
  object, for instance, for which the word \emph{object} was invented,
  are really \emph{objects} and not fleeting and fugitive appearances,
  because they are not only groups of sensations, but groups cemented
  by a constant bond. It is this bond, and this bond alone, which is
  the object in itself, and this bond is a relation }
\citep[\S\S 4 and 6, I underline]{Poincare02a}.
\end{quotation}

The concept of symmetry allows to give a quantitative meaning to the
terms used in this remarkable text: ``Relations'' and
``classification'' have already been discussed in
\S~\ref{subsec:classification}.  ``Translation'' (not taken here to
synonymous of a shift!)  appears to be a passive transformation in the
sense considered in section~\S~\ref{subsec:activepassive}, and its
relation with objectivity has been emphasized
in~\S~\ref{subsec:reprod_object}; ``object'' being considered as an
equivalence class, an ``invariant''. Whereas ``discourse'' (we have
seen above that Houtappel et al. also use  the term ``language'')
corresponds to a choice of convention (but, of course, Poincar\'e is
far from endorsing nominalism, see \S~4 of his \citeyear{Poincare02a}
reflexions), a coordinate chart in differential geometry or more
generally a representative of an equivalence class, that ``breaks'' the
symmetry (or, rather, the invariance).

\section{Comprehension and classification}\label{sec:comprehension}

Poincar\'e's last remark on the importance of classification and its
role in identifying invariants leads us to focus on the second facet of
the concept of symmetry.

Comprehension is an old scholastic notion and it is not by chance that
the same word is used to mean the act of comprehending and the act of
understanding.  The latter requires to find analogies\footnote{See
  \citet[I, 2nd part \textit{Transcendendal logic}, 1st division,
    book II, chap.~II, \S~III.3, \textit{Analogies of
      experience}]{Kant98a} and
  \citet[chap.~IV~\S~49]{Cournot56a}.}, correlations, regularities
between a priori distinct entities or phenomena; these connections are
the most precisely expressed by equivalence relations (incidentally,
relatives constitute the fourth Aristotelian category;
Aristotle \citeyear[\S~IV, pp.~16--17]{Aristotle38a}) which are
synonymous, as we have seen, of classification and therefore of
unification.  The common qualities shared by the elements of one class
are the invariants and allow to define them by intension rather than by extension.
 The duality between a class and its elements
reactivates, to use a terminology with some medieval taste, the old
opposition between universals and individuals.

\subsection{From taxonomy to predictability}

We do not have to probe very deeply to acknowledge the major role of
taxonomy in science. Far from being a purely descriptive tool, sorting
things out is above all relevant when it reveals some objective
properties, as Poincar\'e explains in the last quotation given above.
In biology, for instance, the hierarchical classification initiated by
Carl Linnaeus which were originally based on phenotype only, allowed
to reveal some correlations that could be explained within the general
scheme of the theory of evolution\footnote{More generally, the role
  played by symmetries is of the greatest importance in all
  hierarchy-based philosophies, particularly in Cournot's
  \citeyear[book~IV, chap.~XIV, \S~514]{Cournot1861a}.}. The degree of
similarity between some physical or even behavioural characteristics
of living organisms have been a relevant indicator of the chronology
of the divergence between species. Even before the genetic distance
was even conceived and before phylogenetics, these correlations could be associated (through
correlations of correlations, so to speak) to the history of our
planet (climate fluctuations, atmosphere composition, continental
drifts, etc).

Another example is provided by the appropriate way of classifying
chemical elements which led Mendeleev to some successful
predictions on the properties of elements that were yet unknown in
1869. Without knowing the atomic structure (at an epoch where only few
physicists accepted the existence of atoms), this periodic
classification exhibits some universal characteristics that were not
undermined by the subsequent improvements of the physical theories and
the numerous models of the atoms that came with them (first classical,
like  Kelvin's vortex atoms in 1867 and later on, of course, the
quantum models).

During the second half of the twentieth century, symmetry has
strengthened its classificatory and predictive role.  Originated in
the 1930's by Wigner---who received in 1963 the Nobel price in physics
precisely ``for his contributions to the theory of the atomic nucleus
and the elementary particles, particularly through the discovery and
application of fundamental symmetry principles"---our modern
classification of quantum particles, whether there are supposed to be
elementary or not, relies on the classification of the (linear)
algebraic representations (in the sense defined in
\S~\ref{subsubsec:algrep}) of groups like the Poincar\'e group
(Wigner's philosophical reflections on symmetry are collected in
his \citeyear{Wigner95a} book. For a modern presentation of technical
details see Weinberg, \citeyear[chap.~2]{Weinberg95a}).  Unlike
taxonomy of macroscopic objects, the classification of quantum
particles relies on a very few parameters only, like the mass, the
electric charge, the spin.  The reason is that in quantum physics only
very few observables are compatible with all the others (technically,
those associated with operators that commute with all the others,
which are associated with the so-called superselection rules). Otherwise, an
observed property is generically completely erased by the measurement
of an incompatible observable: quantum particles 
have no history nor
accidents and quantum field theory naturally incorporates the
indistinguishability of the particles belonging to the same
class. Anyway, the hadron classification leading to the prediction of the $\Omega^-$ or the
prediction of the existence of particles like the $W^\pm$, $Z^0$ or
the Higgs boson in
the Standard Model are all built on
symmetry arguments (more specifically on the basis of a symmetry breaking argument). These provide another illustration, like in
Mendeleev's contribution, that the possibility of classification
 is a signature of an objective
reality, the latter being taken in the sense given by Poincar\'e in
the last quotation. In the same time, this classification often reveals a
 substructure (the periodic table reflects the atomic structure, the 
``eightfold way'' to classify the hadrons reflects the quark structure and
maybe the quark and leptons classification reflects a still unknown subworld).

\subsection{Inference and the problem of change}\label{subsec:inference}

Since any time-translation can be considered as a particular case of
symmetry transformation (\S~\ref{subsec:relevantgroups}), the
foresight skills conferred by evolution equations (whether classical
or quantum) appear as the most important aspects of the predictive
power provided by symmetry.  Obviously, this ability of anticipation
covers fields much broader than science. There is no need to solve
differential equations to successfully forecast that the sun will rise
tomorrow morning.  After having closed your eyes, you know with almost
certainty that you will find the words you are currently reading in
the same order once you look again on this page. There is no miracle
here \citep[chap.~4, p.~73]{Putnam75a}, just a consistency for rational
thinking, a \textit{raison d'\^etre}: no thoughts, no language could
be possible if we were unable to discern some ``constant bond'' as
Poincar\'e wrote. These stable properties (at least for a moment long
enough to be noticeable) reveal what we call the existence of things,
building up together what we call the real world (including
ourselves, see for instance Mach, \citeyear[chap.~XXV, \SS~14, 17]{Mach06b}). A macroscopic object like a cherry can be considered as an
equivalence class associated with properties (shape, taste, colour,
size, etc.)  that remain invariant despite the change of light, the
endless adsorption or desorption of molecules (active transformations
in the sense explained in~\S~\ref{subsec:activepassive}) or the
(passive) transformations of the spatial position of the measurer or
of his neural states. As remarked by
\citet[chap.~IV]{Heisenberg58a} or \citet[chap.~5,
  \S~IX]{Popper69a}, this ``problem of change'' (or of becoming) has
already been settled by pre-Athenian philosophers among which
Heraclitus and Parmenides, who proposed the two most extreme solutions
along the whole continuous spectrum of possible answers.

We will come back in section~\ref{sec:invariance} to the notion of
invariance, which constitutes the third facet of symmetry. However, at this
stage, we can usefully notice that symmetry as a predictive tool
provides an intermediate support for the inductive/deductive
inference.  Induction, as a reasoning from a part to a whole is
formally described by the operation~$x\mapsto\sigma(x)$ while
deduction, processing from general to particular instances, offers a
second look at the fourth facet of symmetry, formally described by the
operation~$\sigma \mapsto x$.

\subsection{Stability and abstraction}\label{subsec:stabilityabstraction}
As analysed by Taine 150 years ago,  any intelligent 
being is characterised by his ability of constructing a 
representation of the world \citep[see also][chap.~VII, \S~109]{Cournot56a}:
\begin{quotation}\small{
Thus, every normal
sensation corresponds to some external fact which it transcribes 
with greater or less approximation, and whose internal
\emph{substitute} it is. By this correspondence, internal events agree
with external, and sensations, which are the elements of our
ideas, find themselves naturally and beforehand adjusted to
things, which adjustment will, further on, enable our ideas to
be in conformity with things, and consequently true.
\am{ On the
other hand, we have seen that images are \emph{substitutes} for 
sensations, past, future, or possible, that individual names are
\emph{substitutes} for images and sensations momentarily absent, that
the more simple general names are \emph{substitutes} for images and
impossible sensations, that the more complex general names
are \emph{substitutes} for other names, and so on. It seems then
that nature has undertaken to provide in us representatives
of her events, and has effected her purpose in the most economical way. 
She has provided, first, the sensation which translates 
the fact with more or less precision and delicacy; then,
the surviving sensation capable of indefinite revival, that is to
say, the image, which repeats the sensation, and consequently
translates the fact itself; then, the name, a sensation or image
of a particular kind, which, by virtue of its acquired properties,
represents the general character of many similar facts, and 
replaces the impossible images and sensations which would be
necessary to translate this isolated character. By means of
this correspondence, of this repetition, and this replacement, 
external facts, present, past, future, special, general, simple,
or complex, have their internal representatives, and this mental
representative is always the same internal event, more or
less compounded, repeated, and disguised.}}
\citep[end of book III, chap.~II, \S~V, p.~141\am{-142}]{Taine1872a}.
\end{quotation} 

This correspondence between the world and a mental
representation\footnote{Not surprisingly there is an abundant
  literature that discusses the congruence between objects and beliefs
  since it determines the nature of truth. As old as western
  philosophy itself, this subject is omnipresent in logic, in
  linguistic, in psychology, in jurisprudence, in epistemology etc.
  If we are ready to move the boundaries as suggested in
  \S~\ref{subsec:movingboundaries}, the traditional opposition
  between coherence and correspondence theory of truth should be
  softened \citep[see also the
    discussion in][\S~17, p.~117]{Weyl49a}.}  is necessarily partial not only because a brain,
as a tiny part of the universe, can only encode far less information
than the universe itself \citep{James1880a}, but also for the sake of
the efficiency that enforces its own stability as an organised
structure, as needed by Darwinian competitive selection
(Popper, \citeyear{Popper79a}, \S~2.16;
Dawkins, \citeyear{Dawkins76a}, chap.~4;
Dennett, \citeyear{Dennett96a}, chap.~4). The processing of
information---memorising, learning, anticipating---by any intelligent
entity cannot be done without a discrimination between relevant and
irrelevant parameters \citep[by the way, the
  famous Wisconsin card sorting test allows neuropsychologists to
  quantify the adoption of classification rules and measure its
  flexibility by trial and error, see for instance \S~6.5, p.~412 of][]{Baars/Gage10a}.

Ireneo Funes, the character dreamt by Jorge Luis Borges, 
whose memory was rigorously
infallible,
\begin{quotation}\small{
 was almost incapable of
general, platonic ideas. It was not only difficult for him to
understand that the generic term dog embraced so many unlike specimens
of differing sizes and different forms; he was disturbed by the fact
that a dog at three-fourteen (seen in profile) should have the same
name as the dog at three-fifteen (seen from the front)} 
\citep[Funes, the memorious, p.~114]{Borges42a}.
\end{quotation}

He who cannot forget cannot generalise; he who cannot generalise
cannot think.  Classification appears  therefore as a fundamental
thought process and hence has been the subject of a wide field of
research in cognitive sciences where the synonym ``categorisation'' is
usually preferred (Rosch, \citeyear{Rosch78a}\nocite{Rosch78a},
Smith, \citeyear{Smith95a}\nocite{Smith/Osherson95a};
Murphy, \citeyear{Murphy02a}, chap.~7; Machery, \citeyear{Machery09a},
chap.~6, for recent surveys of this matter).  Abstraction, forging
concepts, using a language, constructing models and eventually
building up theories are variations on the same theme, namely, the
second facet of symmetry. Observation itself appears to be an act of
abstraction---actually the lowest level of abstraction---
inconceivable without entities, the so-called observable entities,
that are indeed concepts \citep[part~2, chap.~IV, \S\S~1,2 and
  chap.~VI, \S~1]{Duhem54a} emerging when pruning the information out
of a bunch of data extracted from a magma of occurrences. Following
\citet[chap.~1, \S~C]{Hanson58a} and talking about theory-laden
observations is therefore awkward, unless we differentiate between multiple
levels of abstraction, because it supposes a theory-independent core to
be loaded and, thus, this leads immediately to a self-contradiction
since this core appears to be an abstract notion in itself. In that
sense, in my criticism of empiricism, I am going further than Hanson and
according to \citet{Sellars56a}\nocite{Feigl/Scriven56a} I would
rather prefer talking about the ``myth of the given''.
\begin{quotation}\small{In sum, facts are facts, and \emph{if it happens that they satisfy
a prediction, this is not an effect of our free
activity.} There is no precise frontier between the fact in the
  rough and the scientific fact; it can only be said that such an
  enunciation of fact is \emph{more crude} or, on the contrary, \emph{more
  scientific} than such another} \citep[\S~3]{Poincare02a}.
\end{quotation} 
De facto, the degree of precision of any classification depends on the
criteria we retain. Fruits can be classified according to the kind
of desserts you can make with or according to the genome of its plant.
Stars can be grouped in constellations or in galaxies 
\citep[chap.~XI, \S~161]{Cournot56a}.  The criteria are always 
moving, are often
arbitrary or contingent and maybe difficult to define without referring
to the category itself: the most efficient way to explain to a child
what a cherry is remains to show him several examples of this fruit,
not to give a definition \citep[\S~4]{Poincare03a}.  This intuitive
approach is all the more natural than brains are instinctively
inclined to think in terms of associations that nurture its creativity
\citep[for instance]{Hadamard54a,Weil60a}.  Unlike computers, brains are
analogue and selectionist rather that digital and instructionist
devices (Popper, \citeyear{Popper78a}, p.~346; Edelman \citeyear{Edelman92a}, specially the critical postscript, 
Dehaene, \citeyear{Dehaene97a}, specially the chap.~9).  The
reluctance of being classified offered by certain elements requires
reflection, testifies the weakness of our comprehension and gives
evidence of the rich variety and of the impermanence of the world. A
thing or a word can come with many variations of significations,
without which there cannot be nuances, ambiguities, metaphors and
even poetry.

But unlike art, rational thinking and, consequently, its most
elaborated form that science represents, tends to sharpen and
perennialise the boundaries of the equivalence classes; it brings out
criteria of truth beyond cultural barriers and socio-historical
context.  This strategy allows to reach provinces far away from our
natural (i.e. adequate from an evolutionary point of view) narrow
county and, as we have seen in the previous section,  to make
predictions \citep[specially chap.~2]{Llinas01a}.  I have already
mentioned (\S~\ref{subsec:hierarchies}) the hierarchy of numbers in
mathematics; it extends the concept of numbers much further than the
primitive mental representations of integers, genetically encoded with
a broad brush in highest mammal brains
\citep{Dehaene97a,Butterworth99a}.  In physics, the both prominent
frameworks of general relativity and quantum theory constitute charts
at scales considerably much wider than our every-day life environment.
If we adopt the adequation I propose between existence and equivalence
classes, the questions of knowing if we invent or rather discover
these provinces, if symmetry unveils or sculptures reality, appear to
be unessential \citep[see also][]{Margolis/Laurence07a}; it is a
matter of taste depending on where we prefer to observe the virtuous
circle 
\citep[for a particularly interesting examination of
  this old subject, namely the ``reality of abstractions'', see][chap.~XI, specially the arguments in \S\S~152 and~155]{Cournot56a}.  In
addition, this orientation also allows to soften the question of the
adequacy between the mathematical world and the physical world: if we
merge these two worlds then the effectiveness of mathematics does not
appear as an unreasonable miracle \citep{Wigner60a} but it is still
true that it is puzzling to see how far this fitness can spread.

\subsection{Different levels of Darwinian selection}

The segmentation of the world in equivalence symmetry classes, and the
drawing of more or less fuzzy boundaries between them (the most
fundamental being the partition in $\mathcal{M}$, $\mathcal{S}$ and
$\mathcal{E}$ as explained in \S\S~\ref{subsec:activepassive} and
\ref{subsec:movingboundaries}), can be seen as a selection process.
Jacques \citet[first note of chap.~III, p.~29]{Hadamard54a}
recalls that the Latin original meaning of intelligence is ``selecting
among'' (inter-legere) and echoes the Poincar\'e's statement:

\begin{quotation}\small{
What, in fact, is mathematical discovery? It does not consist in making
 new combinations with mathematical entities that are already known. 
That can be done by any one, and the combinations that can be so formed
would be infinite in number, and the greater part of them would be absolutely devoid of interest. Discovery consists precisely in not constructing 
useless combinations, but in constructing those that are useful,
 which are an infinitely small minority. Discovery is discernment, 
selection} \citep{Poincare08a}.
\end{quotation}

It is of course Darwin's major contribution to have shown how
differential survival governs the evolution of living species. This
natural selective stabilisation is a (non-linear) mechanism at work in
many other dynamical systems made of different entities whose
populations are susceptible of attenuation and amplification (via
reproduction, replication, combinations, etc.). It would be too long a
digression, and beyond my competence, to examine the subtle
distinctions of how Darwinian selection is implemented in so various
fields as immunology, cognitive sciences, computer science,
sociobiology and culture.  Even in evolutionary biology, the details
of the picture are still objects of scientific controversies among
specialists like Richard Dawkins, Stephen Jay Gould, William
D. Hamilton, Richard Lewontin, John Maynard Smith, Edward Osborne
Wilson, to name some popular contemporary figures among many others
\citep{Segerstrale00a,Shanahan04a}.

With the philosophical bet of building a conception of the 
world in a coherent ontology the stakes are too high for the 
philosophy of physics alone ; it must as well rely on a Darwinian like
naturalization that makes some norms to emerge.
To remain within the scope of the present paper, I just wanted to
point out that the formation of some equivalence classes I denoted
by~$\sigma$ can be seen as a natural process \`a la Darwin not only
when $\sigma$ denotes a living species, a pool of genes or of memes
\citep[elements of material or immaterial
  culture, including knowledge,][chap.~11]{Auger52a,Dawkins76a}, but also if we adopt a
neurobiologist point of view.  Any class $\sigma$, considered as an
object i.e. a part of the physical world (a predator, a cherry, a
pixel, an atom) may be represented in terms of some neural
configuration distributed in several intertwined areas of the brain.
Since more abstract concepts ($\sqrt{2}$, a quantum phase, entropy, a
blue tiger, Prometheus, beauty, freedom, etc.) and, more generally,
ideas are also encoded in brains via the neural material support, the
insurmountable difficulty of delimiting what reality is
 can be understood at the neurophysiological level (the recursive
character appears here through consciousness, i.e. when the model of
the world that characterises intelligence includes a representation of
itself).  Some concepts (a cherry, the Moon, an electron, a number,
Kepler's law, a rotation, yourself) are strongly reinforced and
stabilised by being intricately bound within a large hierarchy,
including observations or sense-data, while some others (a blue tiger,
a tachyon, chess rules, the pumpkin-coach transformation, Zeus),
being less firmly established, have a smaller degree of reality.
As any practitioner of quantum physics knows it 
(even before the modulus of some electronic 
wave-function in quantum corrals has been imaged by scanning measurement), 
the concept of a quantum state  generally refers to a more real entity
that the toy concept of a solid sphere made of gold with a radius of 1~km. 

Recent investigations in neuroscience seem to give a neural basis of
an old suggestion made by \citet{James1880a} and to confirm the
existence of Darwinian (epigenetic) selective mechanisms between
various pre-representations generated by combinatorial games throughout
neuron networks.  Many selectionist processes may be distinguished
that are characterised by transformations of neural states at
different time-scales; compared to the fast dynamics of, say, decision
making, the neural Darwinism involved in learning and memorisation may
concern slower developments of neural circuitry, starting before birth
soon after conception (dendritic and axonal generation or regression,
synaptic reinforcement or inhibition, etc.).  The ``survival of the
fittest" representation would be determined by complex tests of
evaluation, reentrance and reinforcements procedures (driven by
pleasure and pain for instance) that involve the internal dynamics of
the brain as well as its coupling to its external environment
(including other brains).  We still know very few about how this
dynamics is physiologically implemented; however, it shares many common
points with selectionism at another level, namely the evolution of
scientific knowledge. Since the first proposals of an ``evolutionary
epistemology'' by \nocite{Campbell60a}\citet[1960 and their numerous references that
  anticipated the subject even if James is not mentioned]{Campbell59a} and \citet{Popper79a}
\citep[see also][]{Toulmin67a,Toulmin72a,Campbell74a}, this
approach has known many outgrowths
\citep{Radnitzky/Bartley87a,Hull01a}.  For a recent review of the
neural selection introduced by \cite{Changeux+78a, Edelman78a}
called neural Darwinism by the latter (\citeyear{Edelman87a}), see
\citet{Seth/Baars05a}, \citet*{Platek+07a} and
references therein.  For a less specialised presentation see, for instance,
Changeux (\citeyear{Changeux85a,Changeux02a}) or \citet{Edelman06a}.
The emergence of a pattern of successes 
from an ``overwhelming background of failures'' \citep[chap.~7, \S~1, note 5]{Fine96a} in science is therefore explained in an analogous way as 
the formation of stable and viable combinations of genes 
from the overwhelming 
possibilities of deadly mutations (as far as the
cumulative progress of knowledge is concerned, we have not been
able to make up the memetic 
analogue of such an efficient process like sexuality
for generating reliable combinations. However, in these matters,
from the association of ideas, both at the mental level and at the 
human collaboration level,
we are  closer to  the memetic
 equivalent of meiosis than to agamogenesis, see 
however the conservative trends 
described by Toulmin, \citeyear[\S~VI]{Toulmin67a}).

\subsection{Symmetry-assisted research of simplicity}
 
\subsubsection{Simplicity as a general evaluation against underdetermination}

Most obviously, one may fit a given discrete set of registered events with
many different laws \citep[chap.~IV]{Cournot56a}. Still, this so-called
underdetermination, acknowledged under one form or another since
Epicurus (in the \textit{Letter to Pythocles}, see Laertius,
\citeyear[chap.~X, mainly \S\S~86--88]{Laertius25a}) and carefully
examined by \citet{Hume1758a}, still abundantly discussed in
epistemological debates on induction
\citep[chap.~8]{Laudan90a,Psillos99a}. Notwithstanding,
underdetermination appears to be less serious than some philosophers
think it is, provided that we take into account the notion of simplicity as
one of the selection criterion (see Poincar\'e's 1891 citations given above together with
his \citeyear{Poincare1900a} article;
Cournot, \citeyear[chap.~IV, \S~41]{Cournot56a}; 
Mach, \citeyear[\S\S~X.15, XXV.6,7]{Mach06b}
Jeffreys, \citeyear[chap.~IV]{Jeffreys31a};
Popper, \citeyear[chap.~VII]{Popper59a} and \citeyear[chap.~1, point
  (6) of the appendix, chap.~10, \S~XVIII]{Popper69a};
Reichenbach \citeyear[\S~42]{Reichenbach38a}).  Following the same
lines of thought drawn above, the aim of this section is to suggest
how the second facet of symmetry is at work in this Darwinian-like
selective operation.

\begin{quotation}\small{The problem of simplicity is of 
central importance for the
epistemology of the natural sciences. Since the concept of simplicity
appears to be so inaccessible to objective formulation, it has been
attempted to reduce it to that of probability, which has already been
incorporated to a large extent into mathematical thought} 
\citep[\S~21, pp.~155--156]{Weyl49a}. 
\end{quotation}

Here we join back the aesthetical appeal of a symmetry we briefly
recalled in the very beginning of \S~\ref{subsec:fourfacets}.  The
harmony of a theory, the beauty of an argument or the elegance of an
experiment are frequently evoked by scientists
(Fresnel, \citeyear{Fresnel1818a}; Poincar\'e, \citeyear{Poincare08a};
Russell, \citeyear[chap.~IV]{Russell17a}; Hardy, \citeyear{Hardy40a};
Weyl, \citeyear[\S~21]{Weyl49a}; Heisenberg, \citeyear[chap.~XIII]{Heisenberg74a};
Penrose, \citeyear{Penrose74a}; Lipscomb, \citeyear{Lipscomb80a};
Yang, \citeyear{Yang80a}; Dirac, \citeyear{Dirac63a}, \citeyear{Dirac82a};
Chandrasekhar \citeyear{Chandrasekhar87a};
Weinberg, \citeyear[chap.~VI]{Weinberg92a}; Changeux \citeyear{Changeux12a})\nocite{Curtin80a}, some of
them considering it as a constructive principle if not a criterion of
truth.
 For a curious attempt to quantify aesthetics using a, somehow
vague, notion of complexity see \citet{Birkhoff33a}; for a more
serious and interesting study of the role of aesthetics in science see
\citet{McAllister96a} where Haldane's text below is partially
quoted (chapter~7 provides an actualised discussion on simplicity, see
also the diverse contributions on this subject in Zellner et
al., \citeyear{Zellner+01a}).  One reason of this attractiveness is
the gain in simplicity, a parsimony of descriptive means with respect
to the wide range of their scope \citep[see for instance][Appendix I(A), fifth theory, p.~580]{Hamilton1861a}\footnote{Even before
the ``least action'' or other variational principles were acknowledged, a parsimony principle is already present in Galileo's work (``Nature does not act by means of many things when it can do so by means of a few'', \citeyear[2nd day]{Galilee1632a}).  }.\enlargethispage*{\baselineskip}
\begin{quotation}\small{
In scientific thought we adopt the simplest theory which will explain
all the facts under consideration and enable us to predict new facts
of the same kind. The catch in this criterion lies in the word
`simplest'. It is really an aesthetic canon such as we find implicit
in our criticisms of poetry or painting. The layman finds such a law
as $\frac{\partial x}{\partial t} =
\kappa\,\frac{\partial^2x}{\partial y^2}$ much less simple than `it
oozes', of which it is the mathematical statement. The physicist
reverses this judgement, and his statement is certainly the more
fruitful of the two, so far as prediction is concerned. It is,
however, a statement about something very unfamiliar to the plain man,
namely, the rate of change of a rate of change. Now, scientific
aesthetic prefers simple but precise statements about unfamiliar
things to vaguer statements about well-known things. And this
preference is justified by practical success} \citep[Science and
    Theology as Art Forms, p.~227]{Haldane27a}.
\end{quotation}

Broadening the scope of a theory and reducing the number of laws and
primitive concepts are two faces of a single coin. The tension between
these two antagonistic tendencies is the mainspring of the progress of
scientific knowledge. Focusing on just one of those opposite poles
leads to ridiculous positions: on the one hand, if we consider only
very few particular events, we can explain their occurrence by
arbitrary simple or rather simplistic laws and concepts (this is one
of the main characteristic of superstitions and pseudo-sciences) and
the complication comes from the multiplication of such ad hoc
explanations as the number of observations increases (see for instance
Popper, \citeyear{Popper69a}, chap.~1, \S~I point (7) and its appendix
or Worrall, \citeyear{Worrall89a}, p.~114). On the other hand, as
Feynman puts it with his inimitable style 
\citep[vol.~II,\S~25-6]{Feynman+64a}, from any physical law (not necessarily
discovered yet!) labelled by~$l$, once written in the (dimensionless)
form~$Q_l=0$, we can always introduce the physical concept of
``unworldliness'' associated with this law, namely $U_l\DEFt|Q_l|$
where $|\cdot|$ denotes the modulus or the norm, and write the
``equation of everything'' as
\begin{equation}
  U=0
\end{equation}
where the total unworldliness is given by $U=\sum_l U_l$.  Of course
this trick to hide the complexity under the simplest conceivable
equation remains sterile because of the artificial and shallow
interpretation of~$U$. As Feynman explains it, this stratagem is not
motivated by any symmetry arguments: because it does not take into
account any transformation rules, its nature is very different from
the grouping of the scalar quantities~$v^\mu$ into one compact
geometrical object~$v$ (see \S~\ref{subsec:poincare}
p.~\pageref{eq:vectorcoordinates}).

However, since Pythagoras, we can find many attempts based on
aesthetic and symmetric grounds that have failed to provide
correct explanations. Up to his very last years Kepler did not want to
abandon his cosmological model based on Platonic solids though he
acknowledges, in many notes of the 1621 edition of the Mysterium
Cosmographicum, that his beautiful construction is wrong (25 years
after the first edition and after the discovery of the three laws;
there is some kind of irony here since Kepler himself contributed to
undermine the symmetrical appeal of the ``harmony of perfect spheres''
with the first law which manifests the breaking of the rotational
invariance through one individual elliptic orbit). If the absence of
any experimental evidence in the LHC experiments is confirmed, it may
be that supersymmetry should be abandonned, at least in its simplest
form.

\subsubsection{Symmetry as a pruning tool}\label{subsubsec:pruning}

One of the major flaws of the attempts to define physical
complexity/simplicity from algorithmic information theory is that it
relies on information compression without any 
loss \citep[for an upgraded account on this vivid field and for the original
  references]{Delahaye99a,Li/Vitanyi08a}. In fact, to gain
in simplicity we must be allowed to \emph{irreversibly} cut some
superfluous hypothesis or initial conditions with an Ockham's razor.
Moreover, we have seen in \S~\ref{subsec:stabilityabstraction} that
information erasing is essential because classification is primordial,
ubiquitous and necessary\footnote{A parallel point of view is proposed
  by Landauer (see, for instance his \citeyear{Landauer67a} paper and 
  the collection of papers on information erasure edited by 
  Leff and Rex, \citeyear[specially chap.~4]{Leff/Rex03a}) and
  leads to a modest realism different from the one proposed here.
  Landauer adopts a strong ontological difference between mathematical
  entities and reality by denying the existence of~$\pi$, for instance
  \citep[p.~65]{Landauer99a}. The proposal I prefer to defend here is
  that what $\pi$ and a cherry designate differ one from the other by
  their degree of reality on a continuous scale.}.  We cannot avoid
working with some equivalence classes, some of which we nonetheless
take as primitive entities by not considering the distinct properties
that allow to discriminate its elements, ones from the others. By
quotienting a set in equivalence classes, the second facet of symmetry
is more a mean of elimination than a mean of compression. Here we
stick to the original meaning of abstraction, considered as a
synonymous of removal (Cournot \citeyear[chap.~XI, \S\S~147-149]{Cournot56a} discusses several processes of such a
disentanglement).  Algorithmic reduction appears only to be a special
case of labelling each class by one privileged element~$x^*$
(specifically, one of the shortest) and, then, the algorithmic depth
of~$x$ quantifies the transformation~$\ftransfo{T}$ that maps~$x^*$ to
$x=\ftransfo{T}(x^*)$. However, this algorithmic interpretation can
hardly embrace the physical world without substantial new foundations;
because of continuity and randomness, whether they
are fundamental or just convenient, most, if not all, physical
transformations cannot be reduced in a discrete sequence of binary
operations. As far as I know, nobody has been able to provide the algorithmic
complexity or the algorithmic depth of an electron. 

Here we join back the thermodynamical (or Shannon) conception of
information, where entropy measures the loss of information that
occurs when considering a statistical ensemble rather than the
individual systems (or messages, if we are concerned with information
transmission). The latter is described in terms of a huge number of
variables~$\varphi$ (say, the microscopical variables like the
positions and the velocities of classical atoms) while the former is
characterised by few variables~$\Phi$ (say, the average velocity of
the atoms, their total energy, etc.). The art of statistical physics
consists precisely in identifying quantitatively the relevant
variables~$\Phi$ that allow a tractable account of which~$\varphi$'s
are compatible with~$\Phi$ (\textit{reduction}) and how $\Phi$ can be
obtained from~$\varphi$ (\textit{emergence}). We often speak of
macroscopic variables (resp. of a macrostate) when considering $\Phi$
(resp. the ensemble), by contrast to the microscopic variables
(resp. micro or pure state) that refer(s) to one element. This is
often justified because the spatial extension usually allows to
intuitively discriminate both systems (fluid/molecules,
galaxies/stars, heavy nucleus/nucleons, etc.). It may also happen,
when studying dynamical systems, that we may take advantage of a
separation of time scales between a fast evolution of~$\varphi$ and a
slow, secular, evolution of~$\Phi$.  If we want to circumvent the
favouritism of space and/or time, it is preferable to see the 
mapping~$\varphi\mapsto \Phi$ as the erasing (the decimation as 
Kadanoff would say, \citeyear{Kadanoff00a}, chaps.~13 and~14) of irrelevant degrees of
freedom.  This is the quantitative formulation of the
classification~$x\mapsto\sigma$ when the elements~$x$ (resp.~$\sigma$)
can be labelled with the numerical variables~$\varphi$ (resp.~$\Phi$);
the irrelevant parameters being those, discrete and/or continuous,
that allow to label the different transformations~$\ftransfo{T}$ of
$\mathcal{T}$.

One given statistical ensemble contains microstates that differ one
from the other by transformations of~$\varphi$ compatible with the
macroscopic constraints fixed by~$\Phi$.  Building out the effective
model consists in coarse-averaging that is, integrating out, say
within the partition function, the variables that allow to
discriminate the microstates (in quantum statistical physics this
represents the computation of partial traces of the density operator)
and retaining the dependence in~$\Phi$ only. Some of the individual
properties of the particles---typically the ones involved in the
interactions like the charges but not their individual energy for
instance---are unaveraged and remain as such among the~$\Phi$. How
some properties of the effective system depend on such variables
reflects the degree of universality of the construction. It is one of
the greatest achievements of statistical theory to have indeed exhibited
some quantities (the electromagnetic radiation emitted by a black-body,
 the critical exponents) that depends only on the
grossest features of the system (the dimension of space, etc.)  and
neither on chemical composition nor on the precise form of the interactions, as does the critical
temperature of a second order phase transition. Less universal 
but much simpler examples of~$\Phi$ are the pressure of a fluid or the
absorption coefficient~$\alpha$ of a material. 
Both have their value depend on the details of the microscopic interactions but, 
surprisingly, they can be used to characterise a wide
variety of substances whose microscopic structures
strongly differ one from the other (for instance $\alpha$ is involved in the 
Beer-Lambert law to describe the light transmission through a liquid, 
a gas, a plastic, a glass, a diamond). The emergence of 
universality due to the law of large numbers was already noticed by
\cite[chap~IV, p.~191]{Langevin23a} and, before him, by 
\cite[\S~IV]{Poincare11a} who talked in this case
of the insensitivity to large differences in the initial conditions.

More generally, beyond statistical physics, from all the preceding
considerations presented in this section, we can understand that
picking up a small number of relevant variables~$\Phi$ out of a wide
collection of~$\varphi$ remains an indispensable mechanism of any
cognitive process. 

\subsection{Emergence}\label{subsec:emergence}

The last line of thought conveys us into the core of the debate about
reduction/emergence issues \citep{Bedau/Humphreys08a}.  Since,
recently, there have been many discussions
on these matters \citep{Batterman10a,Butterfield10a}, I will not
venture too long into this wide realm 
\citep[see also for instance][]{Castellani02a}.  At the level of abstraction I
adopt in this article, I formally take up the stance that emergence
stands for constructing the variables~$\Phi$ describing~$\sigma$ from
the variables~$\varphi$ describing~$x$.  Reduction, appears to be the
opposite operation $\sigma\mapsto x$ (the fourth facet of symmetry, I
will have to say more about it below). Thus, we have identified the
two dual elementary mechanisms from which are built the hierarchies
that structure the physical world and shape rational thinking.  We
have encountered an example of mathematical emergence in
\S~\ref{subsec:hierarchies}. In \S~\ref{subsubsec:pruning}, I have
sketchily recalled how statistical physics, within the frame of
renormalisation theory, offers what is perhaps the most precise
quantitative illustration of the rearrangement and elimination of the
degrees of freedoms that allow to establish the relation between $x$
and~$\sigma$. In every-day thinking, this is intuitively accomplished
by the categorisation capacity of what we call the common-sense that
make a cherry or a dog to emerge from a bunch of interweaved organic
molecules (or cells if we want to add some intermediate structure).
 
Identifying what properties are relevant or convenient at one level
and establishing quantitatively their connection with the properties
at other levels remains a formidable challenge, the nub of scientific
research I would say.  There is no general method following
well-established instructions to increase our comprehension (again,
our individual minds and the web of our socially interconnected minds are
rather more selectionist than instructionist, more Darwinian than
Lamarkian so to speak).  This is obvious when following the
reductionist direction; however, it is not less true when climbing up the
stairs in the other way.  Digging out the emergent properties by
picking up a small number of relevant variables among a wide
collection can seldom be done explicitly.  Above, I used the word
``art'' to qualify statistical physics not, of course, to deny the
standards of rational thinking which I fully endorse, nor to indicate
a direction of increasing beauty\footnote{As
  \citet[chap.~VI]{Weinberg92a} writes it, general relativity may
  appear more beautiful than Newton's theory though involving far more
  quantities, namely the infinitely many degrees of freedom of the
  gravitational field. In Newtonian theory, the gravitational field
  does not have a proper dynamics that can be made, even partially, independent of the dynamics of matter. }, but precisely to emphasize that there were
no general rule of how to elaborate an effective theory, and even more
to compute (specially when infinities are to be dealt with), the
$\Phi$ from the $\varphi$. The large number of appropriate parameters~$\Phi$
and the difficulties to establish protocols of their measurements mainly
explains why medicine, social sciences, psychology, economics, ecology,
 agriculture, and even the technical aspects of cooking etc. are less
exact (i.e. predictive with less precision)  
than more quantitatively supported branches of science 
like chemistry or physics.
The objects of investigation of the latters
are not only relatively much simpler
but also the statistical fluctuations are much more reduced by the huge 
number of elements in a sample (as far as I know 
it is only in physics that self-averaging quantities can be exhibited).
  Even in every-day rational thinking, the variables~$\Phi$
remain mostly  qualitative rather than quantitative and naive attempts
of their formalisation  often appear as dangerous oversimplifications.

\section{Invariance}\label{sec:invariance}

In the introduction of~\S~\ref{subsec:relevantgroups}, I have shown
how the second facet of symmetry allows to identify classification
with property attribution. I have explained in
\S~\ref{subsec:inference} why these two equal cognitive capacities,
inherent to any kind of measurement (including perception), can be
viewed as a stabilisation.
\begin{quotation}\small{
The general concept of group pre-exists in our minds, at least potentially. It is imposed on us not as a form of our sensitiveness, but as a form of our understanding ;}
 \citep[conclusions]{Poincare1895a}.
\end{quotation}
 A sine qua non condition
of intelligibility is that some physical properties must remain
unchanged under transformations like some space or time translations,
rotations, etc. I precise ``some'' because this stability necessarily
exists in a finite portion of space-time only: The correlated bundle
of attributes that allow to characterise Borges'dog lasts more than
one minute but certainly no more than 30 years. The norm and direction
of the terrestrial acceleration $\vec{g}$ remain approximately uniform
at laboratory scales but change according to where we are on Earth;
moreover, of course, it is only meaningful in the immediate
neighbourhood of the  Earth, which in its turn has a finite space-time
extension.  Some quantum properties like those of the kaon~$K^\pm$
last $10^{-8}\ \mathrm{s}$ while others like the electron charge (more
precisely the fine-structure ``constant'') seem not to vary on
cosmological scales more than one part in $10^6$ (as far as we can
safely measure up to now)\footnote{By the way, while we construct the
  equivalence classes we consider as physical objects, whether we
  prefer to implement or not invariance under some time-translations
  remains a matter of convenience. Both approaches are admissible and
  broadly equivalent in a non-relativistic approach but, of course,
  the time extension cannot be avoided as soon as relativity is taken
  into account since then, simultaneity becomes a relative
  property. The philosophical numerous discussions between
  endurantists and perdurantists seem therefore neither obscure nor very
  interesting once it is illuminated by physics arguments (see, for
  instance Butterfield's \citeyear{Butterfield05a}  point of view and the
  references provided there). }.

It is the greatest merit of Noether's theorems to provide a locally conserved
quantity associated with each independent one-dimensional continuous
family of transformations under which the functional driving the
dynamics (more precisely, the action together with appropriate
boundary conditions) is invariant.  I will not devote more attention
to this specific aspect of invariance since the implications and the
interpretations of Noether's work have already been scrutinised elsewhere
\citep[notably]{KosmannSchwarzbach10a,Brading/Brown04a}.  What I want
to propose in~\S~\ref{subsec:invlaw} is a more general perspective
where invariance, viewed as the third facet of symmetry, concerns not
only formulations of variational principles but, in fact, any law of
science.  Then, in \S\S~\ref{subsec:modest} and \ref{subsec:invmeme},
I will open this interpretation to even broader epistemic
considerations.\enlargethispage*{\baselineskip}
\begin{quotation}\small{This rudimentary tendency toward ``objectification''
reappears in conceptual, in particular mathematical, thought, where it
is developed far beyond its primitive stage. When we determine the
size of an object by \emph{measurement}, it is owing to such
``objectification'' that we succeed in transcending the accidental
limits of our bodily organisation. It enables that elimination of
``anthropomorphic elements'' which is, according to Planck, the proper
task of scientific natural knowledge. To geometrical invariants have
to be added physical and chemical constants. It is in these terms that
we formulate the ``existence'' of physical objects, the objective
properties of things.[\dots] Hering speaks [\dots] the language of the
scientist, i.e., of realism. He assumes the empirical reality of the
objects about which our senses have to inform us. But a critical
analysis of knowledge must go farther. Such an analysis reveals that
the ``possibility of the object'' depends upon the formation of
certain invariants in the flux of sense-impressions, no matter whether
these be invariants of perception or of geometrical thought, or of
physical theory. The \emph{positing}  of something endowed with
    objective existence and nature depends on the formation of
    constants of the kinds mentioned.  It is, then, inadequate to
    describe perception as the mere mirroring in consciousness of the
    objective conditions of things. The truth is that the \emph{search for
    constancy}, the tendency toward certain invariants, constitutes
    a characteristic feature and immanent function of perception.
   This function is as much a condition of perception of objective existence 
   as it is a condition of objective knowledge}
    \citep[end of~\S~III, pp.~20--21]{Cassirer44a}.
\end{quotation} 
Born is also very clear:
\begin{quotation}\small{
I think the idea of invariant is the clue to a rational concept of reality, 
not only in physics but in every aspect of the world. 
\\\indent
[\dots] not every concept  from the domain of scientific constructs 
has the character of a real thing, but only those which are invariant in
regard to the transformation involved. [\dots]
\\\indent This power of the mind to neglect the differences of sense 
impressions and to be aware only of their invariant features seems to me the most impressive fact of our mental structure.[\dots]
\\\indent
Thus we apply analysis to construct what is permanent in the flux of 
phenomena, the invariants. Invariants are the concepts of which
 science speaks in the same way as ordinary language speaks of ``things," 
and which it provides with names as if they were ordinary things} 
\citep{Born53a}.
\end{quotation} 
I perfectly agree except that I would nuance and would prefer to say
that ``not every concept
shares the same degree of reality'' because, as I will recall below,
 invariance of a ``real thing''  always appears to be an approximation.

\subsection{Science laws as quantitative manifestations of invariance}
\label{subsec:invlaw}

Whatever school of thinking one belongs to, whatever status one
attributes to a science law\footnote{Among the overwhelming literature
  on this subject, I should mention van Fraassen's \citeyear{vanFraassen89a}
  influential attempt  to deny laws of nature while, strangely enough,
  keeping the notion of symmetry. If I am not mistaken, what van Fraassen
 criticises  is the dogmatic conceptions of laws, specially when we
forget that their quantitative formulations should come
with their conditions of 
validity.}, it ought to
be uncontroversial that most, if not all, science laws yields to
relations between quantities~$x$ that can always be written
as\footnote{Inequalities, like Heisenberg's or the second law of
  thermodynamics can also be turned to this form with the help of
  Heaviside step function.  The mathematical relation between
  the~$x$'s is of course not unique.  For instance, $E-mc^2=0$
  and~$\exp(1-E/mc^2)-1=0$ are strictly equivalent expressions.
  However, here also, a principle of parsimony should help to select a
  convenient form.}
\begin{equation}\label{eq:Req0}
  \mathscr{R}(x)=0\;.
\end{equation} 
Frequently, these relations are written in the most explicit form
\begin{equation}\label{eq:qQ}
  q=\mathscr{Q}(y)
\end{equation}
by privileging among the~$x$'s some quantities, namely $q$, against
others, namely $y$. I use here two different fonts to discriminate
between mappings, here~$\mathscr{Q}$ and $\mathscr{R}$, and the
physical quantities, here~$x$, $y$ and $q$.  These mathematical
relations (often taken by many to be synonymous of the laws
themselves) have a physical meaning only when they are embedded in a
theory where the quantities~$x$ or $(q,y)$, have been related to
observations, via a model incorporating more or less intermediate
other quantities.  This does not imply that all the~$x$'s should be
uniquely determined by measurements; some of them may be unobservable,
superfluous or irrelevant.  Moreover, when talking about laws,
relations \eqref{eq:Req0} or \eqref{eq:qQ} should not be confused with
definitions. ``All ravens are black'' can represent either a partial
definition of what a raven is or a law if the class of ravens and the
blackness have been defined by other ways.  Both status are mutually
exclusive but remain a matter of choice as we keep these instances in
a coherent network of significations.  History of science is full of
examples where theorems/laws/synthetic judgements have switched to
axioms/definitions/analytic judgements and vice versa.  See, for
instance, the discussion by \citet[\S~17, p.~114]{Weyl49a} on the
status of the electric field~$\vec{E}$ in the equation giving the
electrostatic force~$\vec{F}=e\vec{E}$; or Feynman's 
\citeyear[\S~12-1]{Feynman+64a} explication of the meaning of the
force as it appears in Newton's second law~$\vec{F}=m\vec{a}$.

A transformation~$\ftransfo{T}$ that is performed on the variables~$x$
defines a transformation of~\eqref{eq:Req0} according to the rules
\begin{equation}
  \T{\mathscr{R}}(\T{x})=\mathscr{R}\big(x(\T{x})\big)
\end{equation}
where, in the right hand side, $x$ are considered as functions of the
transformed variables~$\T{x}$. Then, in the second case, the transformed
function~$\T{\mathscr{Q}}$ is defined by
\begin{equation}
  \T{q}=\T{\mathscr{Q}}(\T{y})\;.
\end{equation}
Relation~\eqref{eq:Req0} will be said invariant if the same
relation exists between the transformed quantities~$\T{x}$:
\begin{equation}
  \mathscr{R}(x)=0\qquad\Longleftrightarrow\qquad
  \mathscr{R}(\T{x})=0\;.
\end{equation}
If~$q$ stands for just one scalar quantity with respect to~$\ftransfo{T}$
 i.e. that by definition~$\T{q}=q$ (this not an essential point, the argument
 can be adapted if we were to work with non trivial 
linear or even non-linear algebraic
representations of~$\ftransfo{T}$), then invariance
of~\eqref{eq:qQ} is formally equivalent 
to the equality between the two functions~$\T{\mathscr{Q}}$ and 
$\mathscr{Q}$, i.e. for all $y$ 
\begin{equation}
  \T{\mathscr{Q}}(y)=\mathscr{Q}(y).
\end{equation}

In some circumstances, using the tools developed in catastrophe theory,
 the structural stability of relations
\eqref{eq:Req0} or \eqref{eq:qQ} is reasonably guaranteed (Poston and Stewart, \citeyear{Poston/Stewart78a}; Arnold, \citeyear{Arnold84a}; Demazure, \citeyear{Demazure00a}, to retain three remarkable introductions to this subject). 
Anyway, the
general notion of invariance I want to consider here comes closer to
Poincar\'e's conception that pervades in his 1902 quotation in
\S~\ref{subsec:poincare}.  Let us just mention two simple examples,
involving neither integrals nor derivatives and only scalars, not
tensors, not fields, nor operators: Kepler's third law connecting
the orbital period~$P$ of a planet whose semi-major axis length is
$a$:
\begin{equation} \label{eq:kepler}
  \frac{P^2}{a^3}-\frac{4\pi^2}{GM}=0
\end{equation} 
($M$ is the mass of the Sun, $G$ stands for the gravitational constant); and 
the equation giving the period~$P$ of 
the harmonic oscillations of a mass~$m$ attached to an ideal spring
of strength~$\kappa$:
\begin{equation}\label{eq:oh}
  P=2\pi\sqrt{\frac{m}{\kappa}}\;.
\end{equation} 
First, these equations manifest some invariance with respect to the
usual transformations that constitute the Galilean group and that can
be both interpreted passively or actively. Indeed,
equation~\eqref{eq:kepler} remains as correct now as it was 350 years
ago. Equation~\eqref{eq:oh} is valid in any laboratory on Earth, even
if embarked on a boat  with constant speed.
However, the \textit{raison d'\^etre} of a law is an even broader notion of
invariance: if we actively transform the mass-spring system by varying
the mass $m\mapsto\T{m}$ and/or substituting the spring by a different
one $\kappa\mapsto\T{\kappa}$, we may wonder if the transformed period
of the oscillator will still be given by the same law
 \begin{equation}
  \T{P}=2\pi\sqrt{\frac{\T{m}}{\T{\kappa}}}\;.
\end{equation} 
Kepler's third law is valuable not only because it is still accurate
for Mars since Kepler's century, but also because we can keep the
same relation if we jump from Mars to Venus (then, formally, $a$ and
$P$ are transformed into $\T{a}$ and $\T{P}$ but neither~$M$ nor $G$
are altered) or if we change the centre of attraction by considering
the orbital periods of the Galilean moons around Jupiter (now,
formally $M\mapsto\T{M}$ as well). Even more, $G$ is qualified as a
genuine universal constant precisely because in a given system of
units (that encapsulates connections with other quantities
conventionally considered as standard ones), its value remains
unchanged when performing the previous transformations.

These considerations can be transposed without difficulties to any
field mature enough to be described by reasonably quantitative laws;
for instance to Mendel's statistical laws of heredity, to equations
governing some Darwinian dynamics or to Kleiber metabolic law (for refinements on this latter subject, 
see for instance Brown and West, \citeyear{Brown/West00a}). 
As we recalled in the introduction, it
is in modern quantum theories of fields and in relativity that
invariance has been first set up as a constructive principle. What is
postulated is the a priori invariance of $q$ (a Lagrangian density, an
action, a partition function, a transition amplitude, etc.)  with
respect to some transformations or, in other words, the independence
of $q$ with respect to some quantities among the~$y$'s that appear to
be superfluous (the absolute spatial position, the gauge fields,
etc.). It happens that these symmetry principles may impose, on the
choice of~$\mathscr{Q}$, constraints stringent enough to select the
models drastically and compel specific dynamics on the interactions
(gauge symmetries dictate the properties of the photon or the graviton
for instance). This line of pursuit yields to a feeling of deep
inevitability \citep[chap.~VI]{Weinberg92a}, despite the criticisms
of ``\textit{the
reason which accounts for} uniformity in nature'' made by van Fraassen 
(\citeyear[chap.~II, \S~2]{vanFraassen89a}).

\subsection{Modest truth and humble universality}\label{subsec:modest}

I hasten to add that relations like \eqref{eq:Req0} or \eqref{eq:qQ}
remain always conditionally true. Ideally, in mathematics, theorems
should be formulated with all their hypothesis. In the same way, when
faced with empirical verification and with numerical tests, laws should
come with some error bounds, whether of statistical or systematical
origin, even if their conditions of validity can never all be
explicited \citet[\S~II.B]{Ullmo69a}. In other words, 
as far as predictivity is at stake, we should include an evaluation
of expected risk or loss.
\begin{quotation}\small{If we look at any particular law, we may be certain in advance
  that it can only be approximative. It is, in fact, deduced from
  experimental verifications, and these verifications were and could
  be only approximate. We should always expect that more precise
  measurements will oblige us to add new terms to our formulas this
  is what has happened, for instance, in the case of Marriotte's
  law.\\ \am{\indent Moreover the statement of any law is necessarily
  incomplete. This enunciation should comprise the enumeration of
  \emph{all} the antecedents in virtue of which a given consequent can
  happen. I should first describe \emph{all} the conditions of the
  experiment to be made and the law would then be stated: If all the
  conditions are fulfilled, the phenomenon will happen.\\ \indent
  But we shall be sure of not having forgotten \emph{any} of these
  conditions only when we shall have described the state of the
  entire universe at the instant $t$; all the parts of this universe
  may, in fact, exercise an influence more or less great on the
  phenomenon which must happen at the instant $t + \mathrm{d}t$.}\\ \indent 
   \indent Then as one
  can never be certain of not having forgotten some essential
  condition, it can not be said: If such and such conditions are
  realized, such a phenomenon will occur; it can only be said: If such
  and such conditions are realized, it is probable that such a
  phenomenon will occur, very nearly} \citep[\S~5]{Poincare02a}.
\end{quotation} 
Indeed (see \S~\ref{subsec:movingboundaries}), any modelisation
requires an abstraction of the relevant quantities describing the
system~$\mathcal{S}$ from the irrelevant degrees of freedom that are
swept away in the environment~$\mathcal{E}$.  The mass~$m$ of the
planet does not appear in Kepler's third law but this is true up to
terms of order~$m/M$. The latter may be not negligible any longer if
we want to describe a binary star. If damping is at issue, among the
infinitely many external parameters that have not been considered in the
spring-mass model, the air viscosity must be incorporated in the right
hand side of \eqref{eq:oh} together with a reinterpretation of~$P$ as
a pseudo-period. For transformations such that~$\T{m}$ is too large,
we may also quit the linear regime; then, the period will depend on
the initial amplitude (but still not on the position of the moons of
Jupiter!).

One famous criticism against a realistic conception of science is the
pessimistic (meta-)induction according to which the falsity of the
past scientific theories should lead us to conclude that present
theories are not reliable either \citep{Putnam78a,Laudan81a,Laudan84a}.
However, this argument is fallacious \citep[another reason
  was recently given by][]{Lewis01a} because it presupposes for science a too
ambitious, unreachable and eventually meaningless goal.  \am{Following
Z\'enon, the Renaissance protagonist of Marguerite Yourcenar's novel
\textit{The Abyss},
\begin{quotation}\small{
I have refrained from making an idol of truth, 
preferring to leave to it its more modest name of exactitude} 
\cite[p.~123, A conversation in Innsbruck]{Yourcenar76a}.
\end{quotation}}
Indeed, realism should remain ``modest'' \citep{Bricmont/Sokal01a}
since the reduction of information requested by any modelisation
implies that the truth of a statement, of a prediction, can be
conditional only.  The most we can require from a science law is to come
with the knowledge of its limitation and hopefully with a quantitative
control of the errors; but, of course, this can be achieved once the
theory is mature enough and/or embedded in a wider one as a so-called
effective theory. If we are to use a softer meta-induction argument
(qualifying it as pessimistic or optimistic is a litmus test to
determine whether you think that the science cup is half empty or half
full), if we follow the lessons of history of science and
consider that there cannot be but effective theories. The ``Dreams of
a final theory'' \citep{Weinberg92a} remain an act of faith not far
from the idealistic, therefore unrealistic, desire of Platonic 
truth \citep[for a fair and recent clarification on the subject see][]{Castellani02a}.
Nevertheless we can argue, for instance, that Newtonian mechanics
remains universally true for macroscopic objects with small enough
masses and velocities. Albeit we know relativity and quantum
physics, computation of the location and the dates of eclipses will
still be correctly done within the old classical framework; the main
improvements on the precision of the predictions and retrodictions of
eclipses are expected to come from a refinement of the planetary
models (possibly with post-newtonian amendments but certainly not with
full general relativity and even less quantum corrections) and are
known to be limited in time by the chaotic character of the many-body
(classical) problem.  In that sense, Pythagoras' theorem or Kepler's
law have not been undermined by the development of non-euclidean
geometry, general relativity or quantum physics.

\enlargethispage*{3\baselineskip}
\begin{quotation}\small{\am{The thing essential is that there are points on which 
all those acquainted with the experiments made can reach accord.\\
\indent The question is to know whether this accord will be durable
  and whether it will persist for our successors. It may be asked
  whether the unions that the science of today makes will be confirmed
  by the science of tomorrow. To affirm that it will be so we can not
  invoke any \emph{a priori} reason; but this is a question of fact,
  and science has already lived long enough for us to be able to find
  out by asking its history whether the edifices it builds stand the
  test of time, or whether they are only ephemeral constructions.\\ }
  \indent Now what do we see? At the first blush it
  seems to us that the theories last only a day and that ruins upon
  ruins accumulate. Today the theories are born, tomorrow they are
  the fashion, the day after tomorrow they are classic, the fourth
  day they are superannuated, and the fifth they are forgotten. But if
  we look more closely, we see that what thus succumb are the
  theories, properly so called, those which pretend to teach us what
  things are. But there is in them something which usually
  survives. If one of them has taught us a true relation, this
  relation is definitively acquired, and it will be found again under
  a new disguise in the other theories which will successively come to
  reign in place of the old  } \citep[\S~6]{Poincare02a}\footnote{See also Poincar\'e's quotation given by \citet[p.~103]{Worrall89a}.}.
\end{quotation} 

\am{For the same essential reasons that lead us to estimate the degree of
reality with a continuous balance, I prefer not to consider truth to
be a binary quantity. I will not follow \citeauthor[specially
  chap.~IX, X, XII, XIII]{Kuhn70a} neither \citeauthor{Feyerabend62a}
on their emblematic fields: talking about incommensurability between
two theories is often  an overstatement and I would keep the term to
qualify logically contradictory assertions or dogmatic ideologies only
(to play with the words, recall that etymologically `symmetry' is the
Greek form of the latin `commensurable').  Thus the notion of
scientific revolution is just a convenient tool, at best, to simplify
the vivid evolutionary  river basin of science made of bifurcations as well 
as fusions of many streams. As usual, there is no
fundamental cleavage but a (seemingly continuous) gradation that
separates entities different in degree but not in nature.}

Science laws including their domain of validity appear to be the far
most stable pieces of knowledge; they give evidence for reality or,
rather, these hard core knowledge have the highest degree of
reality. This is why not only any scientific knowledge must be
falsifiable in the Popperian sense, but it definitely must be
falsified.  It is only when a law fails that we can estimate its
domain of validity; then, we can expect to get a broader law that will
encapsulate, protect and therefore perpetuate the first one.

\subsection{Epistemic invariance}\label{subsec:invmeme}

Extending the notion of symmetry transformations outside the usual
corral of space-time transformations gives us a bird's-eye view of
what science laws represent.  In particular, the perspective I
proposed above allows to reinforce the connection between invariance
and objectivity (see also Nozick's \citeyear{Nozick01a} general
reflexions on this point where a non-binary degree of truth and objectiveness
is proposed).  Of course we have
preserved elements of human culture much older than Pythagoras'
theorem, Archimedes' or Kepler's laws but, unlike science laws
\citep[the \textit{intermezzo}]{Sokal/Bricmont98a}, most are crucially
dependent on their contextual signification and appear to be fragments
partially detached from their contemporary cultural background; we
appreciate their value with criteria that are now completely
different from the ones that were in vogue at those ancient times. It
is precisely a cornerstone of the structuralism method applied to
various fields of human science (including linguistic and
anthropology) to find out structural invariants that are shared by the
whole of mankind (and even further, the functional invariants that are
shared by substrates of variable structure, likewise the brain
functions in regards to the neuroanatomical organisation, see
Changeux, \citeyear[\S~VI.7]{Changeux02a}):
\begin{quotation}\small{
Any classification is superior to chaos and even a 
classification at the level of sensible properties is a step 
towards rational ordering. It is legitimate, in classifying
fruits into relatively heavy and relatively light, to begin by
separating the apples from the pears even though shape,
colour and taste are unconnected with weight and volume.
This is because the larger apples are easier to distinguish
from the smaller if the apples are not still mixed with fruit
of different features. This example already shows that
classification has its advantages even at the level of aesthetic perception
}\citep[chap.~1]{LeviStrauss68a}.
\end{quotation}

At a philosophical level, as suggested by Weyl in the conclusion of
his philosophical book, despite the apparently irreducible divisions
between numerous philosophical schools, we can identify some common
denominators (the invariants) that reveal some sort of unity beyond
what is often---but of course not always---merely a semantic
variation, a matter of subjective preference or a pure intellectual
game.
\begin{quotation}\small{
The more I look into the philosophical literature the more I am
impressed with the general agreement regarding the most essential
insights of natural philosophy as it is found among all those who
approach the problems seriously and with a free and independent mind
rather than in the light of traditional schemes---or if not agreement
then at least a common direction in their development. Whether
one talks about space in the language of phenomenology like Husserl
or `physiologically' like Helmholtz is less important, in view of their
substantial concordance, than it appears to the `standpoint philosophers' 
who swear by set formulae
}\citep[end of \S~23.D, p.~216]{Weyl49a}.
\end{quotation}
I hope to have shown that symmetry is actually a good candidate for
being such an invariant as far as rationalists are concerned, among
whom, at least, Poincar\'e's heirs like \citet{Worrall89a}
\citep[see also more recent approaches 
like the one proposed by][]{Esfeld06a}\nocite{Auletta06a} and
his successors---who belong to the multiple subdivisions of the
structural realist branch \citep{Bokulich/Bokulich10a}---but, maybe, also
the supporters of Fine's natural ontological 
attitude \citep[chap.~7]{Fine96a} or even those who prefer to endorse 
Van Fraassen's constructive empiricism. 

\am{
It would certainly be too presumptuous to paraphrase
Dawkins'expression \citeyear[chap.~3]{Dawkins76a} and consider the
scientific laws to be ``immortal'' meme complexes; however,  it is sure that
they constitute the far most stable ones. History shows that they have
been able to resist the most dramatic memocides (like the multiple
destructions of the library of Alexandria) or the most insidious
eumemisms (like the selective copy of texts during the western Middle Ages
that almost completely stemmed the flow of greek materialist
philosophy initiated by the ancient atomists Leucippus and Democritus,
then adapted by Epicurus). Recall,}

\am{
\begin{quotation}\small{
If, in some cataclysm, all of scientific knowledge were to be
destroyed, and only one sentence passed on to the next generations of
creatures, what statement would contain the most information in the
fewest words? I believe it is the atomic hypothesis (or the atomic
fact, or whatever you wish to call it) that \emph{all things are made of
atoms---little particles that move around in perpetual motion,
attracting each other when they are a little distance apart, but
repelling upon being squeezed into one another.} In that one sentence,
you will see, there is an enormous amount of information about the
world, if just a little imagination and thinking are applied}
   \cite[vol I. \S~1-2]{Feynman+64a}.
\end{quotation}}

\section{Projection}\label{sec:projection}

\subsection{Invariance breaking}\label{subsec:invbreak}

We have already encountered the fourth facet of symmetry in various places.

From an epistemic point of view, this is precisely the reductionist
strategy.  The choice of one element in an equivalence class
constitutes the basic operation of decomposing a class in its
constituents: real numbers enclose rational numbers, the class of
fruits comprises cherries, the class of dogs includes Borges'dog,
atoms contain electrons, one-phonon states are built from condensed
atomic states, etc.  Together with the elaboration of effective
models and the identification of emergent properties discussed in
\S~\ref{subsec:emergence}, reduction ties the hierarchical web of
interlaced concepts.

From a logical point of view, we talk about deduction in the sense
that we infer some properties on an~$x$ as soon as it is identified as
an element of class~$\sigma$ (e.g., ``Hume is mortal because humans
are''). In that sense, $x$ may be considered as a prototype (I prefer
to use this word, following Rosch, \citeyear{Rosch78a},
 rather than the too pejorative ``stereotype''  or the
 too much connoted by
idealistic flavour ``archetype'') of the class and a systematic choice of one
representative in each class can be used to label the classes, specially
if a natural (canonical) rule can be proposed (for instance, rational
numbers are represented and can even safely be identified to a
fraction of coprime integers). Here, when talking about the
representation of~$\sigma$ by~$x$, both acceptions b) and c) in
footnote~\ref{fn:representation} fit in this scheme.

There is an alternative way to interpret the projection
operation~$\sigma\mapsto x$.  The identification of one element~$x$
among all the others is possible only if we are able to attribute
properties to $x$ that are not characteristic of~$\sigma$ but, rather,
whose values allow to discriminate its elements. The differences are
encapsulated in the set~$\mathcal{T}$ of transformations connecting
two elements. Because of these distinctive properties, the global
invariance of~$\sigma$ under~$\mathcal{T}$ is broken since
$x\neq\T{x}$ in general. As acknowledged by Pierre Curie in a famous
paper, asymmetry is essential to characterise the physical phenomena:
\begin{quotation}\small{What is necessary is that certain symmetry elements
do not exist\footnote{I translate literally the 
original French. Rosen and Copi\'e's \citeyear{Rosen82a} English translation 
interprets the words ``\textit{n'existent pas}'' in a more appropriate 
and careful way
and propose ``are missing'' instead of ``do not exist''.}.} Asymmetry is what creates the phenomenon
\citep[\S~IV, p.~400]{Curie1894a}.
\end{quotation}
For more modern variations on this theme, see the interesting essay 
by \citet[On Broken symmetries. pp.~99-114]{Morrison95a}.

In practise, the manipulation of concepts often requires the use of one of
its representative. For instance, in theoretical physics computing
scalar products requires the choice of a representation of each vector by
introducing a specific basis. In differential geometry a coordinate
chart is most often required to proceed with numerical
computations. When making measurements, some frame, gauge and units
must be used and, more generally, a measurer $\mathcal{M}$ maps some
part of the universe that includes the physical system at issue (see
\S~\ref{subsec:activepassive} and figure
\ref{fig:transfo_passive_active}). Let us hand over to Weyl
\begin{quotation}\small{
A typical example of this is furnished by a body whose solid shape
constitutes itself as the common source of its various perspective
views
}\citep[\S~17, p.~113]{Weyl49a}.
\end{quotation}
and Born, who uses an analogy with 
 the different elliptical shapes of the projected shadows 
of one circular cardboard  \citep[this
 simple but acute example is already in][chap.~XIII, \S~198]{Cournot56a},
\begin{quotation}\small{
This root of the matter is a very simple logical distinction, which
seems to be obvious to anybody not biased by a solipsistic
metaphysics; namely this: that often a measurable quantity is not a
property of a thing, but a property  of its relation to other
things. [\dots] Most measurements in physics are not directly
concerned with the thing which interest us, but with some kind of
projection, this word taken in the widest possible sense. The
expression coordinate or component can also be so used. The projection
(the shadow in our example) is defined in relation to a system of
reference (the walls, on which the shadow may be thrown)
  }\citep[p.~143]{Born53a}.
\end{quotation}
In any case some superfluous variables are introduced that mask the
objectivity by breaking the invariance of the equivalence classes.
This is, I guess, the main source of confusion that obscures some
ontological or epistemological controversies. One should not confuse
the invariance/objectivity/universality of the class with the
asymmetry/subjectivity/conventional choice of one of its
representation. As Debs and Redhead write
\enlargethispage*{\baselineskip}
\begin{quotation}\small{[\dots] the claim that these 
representations are inescapably based on conventional choices has been
taken by many as a denial of their objectivity. As such, objectivity
and conventionality in representation have often been framed as polar
opposites.[\dots]\\\indent
 [However], a view of science that emphasizes the
role of conventional choice need not be in conflict with a realist
account of representation that allows for objectivity. One may
maintain a cultural view of science and still be committed, as most
realists are, to the existence of a single real ontology that humans
inhabit }\citep[Introduction, pp.~3 and 4]{Debs/Redhead07a}.
\end{quotation}
 
I do not know if Born, when writing the above text, had also in mind
Plato's famous allegory of the cave; however Weyl, in pursuing his
philosophical reflections, clearly endeavours to reconcile realism
with idealism (p.~117).  Then, in almost exactly the same terms we
discussed in \S~\ref{subsec:poincare}, the distinction between the
vector~$v$ (``analogue of the objects in the real world'') and its
components~$v^\mu$ (``analogues of the subjective phenomena") obtained
by projection on a specific coordinate basis (``analogues of real
observers''), Weyl concludes
 \begin{quotation}\small{
Hence the model is the world of my phenomena and the absolute basis
is that distinguished observer `I' who claims that all phenomena are
as they appear to him: on this level, object, observer and appearance
all belong to the same world of phenomena, linked however by relations among which we can distinguish the `objective' or invariant ones. Real 
observer and real object, I, thou, and the external world
arise, so to speak, in unison and correlation with one another by 
subjecting the sphere of `algebraic appearances' to the viewpoint of
invariance. [\dots] The analogy renders the fact
 readily intelligible that the unique `I' of
pure consciousness, the source of meaning, appears under the viewpoint
of objectivity as but a single subject among many of its kind
}\citep[\S~17, pp.~123--124]{Weyl49a}.
\end{quotation}

\subsection{Three open problems}\label{subsec:problems}

As far as contemporary physics is concerned, there exists of course an
enormous literature on the subject of symmetry breaking.  In addition,
part~III of Brading and Castellani's \citeyear{Brading/Castellani03a}
compilation explores interesting
epistemological and philosophical implications of this
concept. I will therefore not say more about these matters. Before we
reach the conclusion, I would like to end this section by listing
three open problems, that I consider to be crucial and more scientific
than metaphysical (I believe that any kind of dualist approach cannot
provide a satisfactory solution but just shifts the emphasis of the
problems). All three seem pertained to the fourth facet of symmetry.

($\mathrm{P_Q}$) \textit{The problem of quantum measurement:} How does
the interaction between a system~$\mathcal{S}$ and a
measurer~$\mathcal{M}$ selects just one observed property or one
observed event among the class of observable ones that a quantum state
encompasses ?  A satisfactory answer must come with a consistent and
unified description of both~$\mathcal{S}$, $\mathcal{M}$ and
possibly~$\mathcal{E}$ unlike the orthodox contemporary quantum theory
which still attributes classical properties to $\mathcal{M}$.
 
($\mathrm{P_N}$) \textit{The problem of now:} In physics, all the
points of the world line of a system are treated on an equal footing.
In the space-time zone where intelligence has emerged, what does
select a special thin slice of time we call present ?

($\mathrm{P_S}$) \textit{The problem of self:} Among all the conscious
minds, what does favour the special one you call your self ?

As Born recalls about a quantum measurement
\begin{quotation}\small{
Expressed in mathematical terms the word projection is 
perfectly justified since the main operation is a direct 
generalisation of the geometrical act of projecting, only in a 
space of many, often infinitely many, dimensions} \citep{Born53a}.
\end{quotation}

Some connections between these problems have been proposed. If
Darwinian selection has not shaped our brains like a digital device,
it has neither prepared us for a natural apprehension of the world at
small scales. After one century of controversy on the quantum
measurement issues, despite the recent progress on the understanding of the role played by environment 
(see note~\ref{fn:measurement}), we still do not know if we will be able to obtain,
within a consistent theory, the reduction of a (possibly diagonalised
by decoherence) density operator to a pure state or if this problem is
irreducible any further because of our intrinsic limitations. This
last line of pursuit follows a tradition initiated by Bohr and later
endorsed  by \citet[part~III]{Wigner95a}  and now
defended by \citet{Penrose97a}, who proposed to relate
$\mathrm{P_Q}$ to $\mathrm{P_S}$. It is worth noting that another approach of $\mathrm{P_S}$ was proposed by \citet[Epilogue]{Schrodinger44a}.

By the way, if we consider $\mathrm{P_S}$ the opposite way by
reversing its terms, viz. if we ask how we can infer the existence of
other selves, we just face the old problem of other minds
\citep{Avramides01a}. Nevertheless, I find the problem of self much
more interesting than the problem of other minds because the latter
finds naturally its answer in the necessity of escaping from a sterile
kind of noological solipsism.

Even though quantum theory of fields offers the most brilliant and
convenient way to reconcile special relativity and quantum theory, it
leaves unchanged the problem of measurement. There seems to be a deep
incompatibility between objective probabilities and a four-dimensional
space-time. According to Everett \citeyear{Everett57a}'s multiworld
interpretation, a bifurcation between universe branches occurs at each
measurement. Had this attempt come with a proposal that clarifies the
special status of measurement against pure quantum events, it would
have ruled $\mathrm{P_Q}$ out but at the cost of reinforcing the
importance of $\mathrm{P_N}$ and especially 
$\mathrm{P_S}$\footnote{It is noteworthy that to try to 
solve another old problem of choice, namely the problem of free will,
Boussinesq has proposed in \citeyear{Boussinesq/SaintVenant1877a}
 to connect it with the existence of bifurcations
of the solutions of dynamical equations, attributing the choice of the branch not 
to an extreme sensitivity to ``a very small change in the initial conditions''
\citep[chap.~IV, p.~76, see also p.~68]{Poincare59a}  but to 
an external ``guiding principle'' (\textit{principe directeur}) that 
establishes a fundamental  clear-cut between inanimate and animate 
beings. 
If, on the contrary, we want to remain self-consistent within a unique 
materialistic continuum, the ``guiding principles'' that break the symmetry
must definitely remain internal. 
 }.

Many physical and philosophical discussions have been devoted to the
arrow of time, but relatively few concern attempts to physically
define the present \citep{Hartle05a}.  Maybe questions $\mathrm{P_N}$
and $\mathrm{P_S}$ are two aspects of the same question, the first
being formulated in physical terms while the other in
neuropsychological terms.

A signature of a final theory should precisely be free of
parameters with contingent value. Those who believe in
the existence of such a theory should therefore add a fourth problem, possibly
connected to the others with the help of anthropic selectionist
arguments (see for instance Bostrom, \citeyear{Bostrom02a} and references therein), namely,

($\mathrm{P_C}$) \textit{The problem of constants :} How are 
the dimensionless constants like the coupling constants at low energy
scales, some cosmological characteristics or even the dimension of
space-time selected?

However, if we deny such an ultimate goal of getting rid of
contingency, the importance of $\mathrm{P_C}$ is weakened because it is
the usual statement of how to digest an effective theory into the
stomach of a wider one.

\section{Conclusion}

Throughout the twentieth century, the notion of symmetry has acquired 
extraordinary scope and depth in mathematics and, especially, in physics.  
In this paper
I have proposed to recognise in this wide domain four intricately
bound clusters each of them being scrutinised into some fine
structure.  First, this decomposition allowed us to examine closely
the multiple different roles symmetry plays in many places in
physics. Second, I have tried to unveil some relations with other
disciplines like neurobiology, epistemology, cognitive sciences and,
not least, philosophy. These excursions and the connections
they reveal, that should gain to be
investigated more thoroughly, offer, I hope an intellectually rewarding
transversal journey.

\am{To put in a shell many themes we encountered, I could not find a
better shell than Val\'ery's; I cannot resist the temptation of
offering it as the last words:}

\am{\begin{quotation}\small{Like a pure sound or a melodic system of pure sounds in the midst of
noises, so a \emph{crystal}, a \emph{flower}, a \emph{sea shell}
 stand out from the common
disorder of perceptible things. For us they are privileged objects,
more intelligible to the view, although more mysterious upon
reflection, than all those which we see indiscriminately. They present
us with a strange union of ideas: order and fantasy, invention and
necessity, law and exception. In their appearance we find a kind of
\emph{intention} and \emph{action} that seem to have fashioned
them rather as man might have done, but as the same time we find
evidence of methods forbidden and inaccessible to us. We can imitate
these singular forms; our hands can cut a prism, fashion an imitation
flower, turn or model a shell; we are even able to express their
characteristics of symmetry in a formula, or represent them quite
accurately in a geometric construction. Up to this point we can share
with ``nature'': we can endow her with designs, a sort of mathematics,
a certain taste and imagination that are not infinitely different from
ours; but then, after we have endowed her with all the human qualities
she needs to make herself understood by human beings, she displays all
the inhuman qualities needed to disconcert us\dots}\cite{Valery64a}.
\end{quotation}}

\bigskip
\textbf{Acknowledgements:} It is a pleasure to thanks X. Bekaert, 
J. Le Deunff, E. Lesigne, K. Morand, K. Noui 
and L. Villain for precious discussions that allowed to 
selectively improve the formulation of some ideas presented here.
 I am also very
grateful towards N. Mohammedi for his careful corrections.

%\bibliography{/home/mouchet/tex/localtexmf/bibtex/bst/mrabbrev,/home/mouchet/tex/localtexmf/bibtex/bst/bibliographie}
%\bibliographystyle{elsarticle-harv}

\end{document}